\documentclass[twocolumn,prc,amsmath,amssymb,aps,showpacs,superscriptaddress,nofootinbib]{revtex4-1}
\usepackage{graphicx}
\usepackage{dcolumn}
\usepackage{bm}
\usepackage{}[multicol]
\usepackage{lineno}

\newcommand{\cerenkov}{\v{C}erenkov}
\newcommand{\kk}{($K^{-}$,$K^{+}$)}
\newcommand{\eek}{($e,e^{\prime}K^{+}$)}

\newcommand*{\KYOTO}{Department of Physics, Kyoto University, Kyoto 606-8502, Japan}

\begin{document}


\title{Development of Water {\cerenkov} Detector for On-line Proton Rejection in 
$\Xi^{-}$ Hypernuclear Spectroscopy via the ($K^{-},K^{+}$) Reaction}

\author{T.~Gogami}\thanks{Corresponding author\\ Email: gogami@scphys.kyoto-u.ac.jp (T.~Gogami)}
\affiliation{\KYOTO}
\author{N.~Amano}
\affiliation{\KYOTO}
\author{S.~Kanatsuki}
\affiliation{\KYOTO}
\author{T.~Nagae}
\affiliation{\KYOTO}
\author{K.~Takenaka}
\affiliation{\KYOTO}






\begin{abstract}
  The missing mass spectroscopy of $\Xi^{-}$ hypernuclei with 
  the {\kk} reaction is planned to be 
  performed at the J-PARC K1.8 beam line 
  by using a new magnetic spectrometer, Strangeness $-2$ Spectrometer (S-2S).
  A {\cerenkov} detector with a radiation medium of pure water 
  (refractive index of 1.33) is designed to be 
  used for on-line proton rejection 
  for a momentum range of 1.2 to 1.6~GeV/$c$ in S-2S.
  Prototype water {\cerenkov} detectors were
  developed and tested with positron beams and 
  cosmic rays to estimate their proton-rejection capability. 
  We achieved an average number of photoelectrons 
  of greater than 200 with the latest prototype for cosmic rays,
  which was stable during an expected beam time of one month.
  The performance of the prototype in the cosmic-ray test was well reproduced 
  with a Monte Carlo simulation in which some input parameters 
  were adjusted. 
  Based on the Monte Carlo simulation, we expect to achieve 
  $>90\%$ proton-rejection efficiency while maintaining $>95\%$ $K^{+}$ survival ratio
  in the whole S-2S acceptance.
  The performance satisfies the requirements 
  to conduct the spectroscopic study of 
  $\Xi^{-}$ hypernuclei at J-PARC.
\end{abstract}

\maketitle



\section{Introduction}
\begin{figure}[!htbp]
  \begin{center}
    \includegraphics[width=8.0cm]{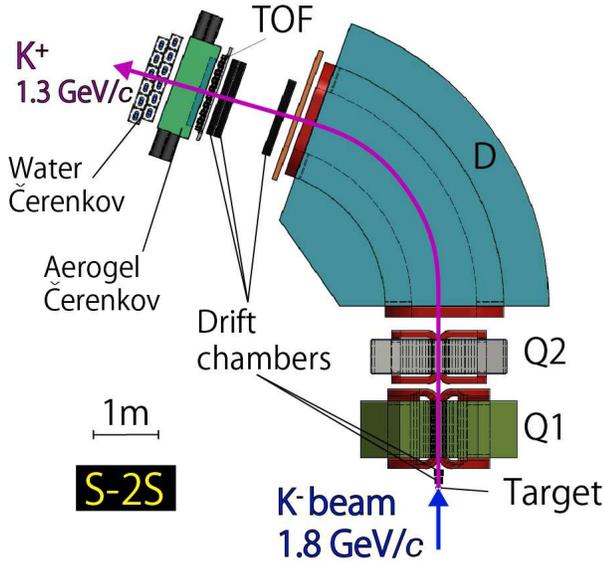}
    \caption{Schematic drawing of the J-PARC E05 experimental setup.}
    \label{fig:s2s_setup}
  \end{center}
\end{figure}

\begin{figure}[!htbp]
  \begin{center}
    \includegraphics[width=8.6cm]{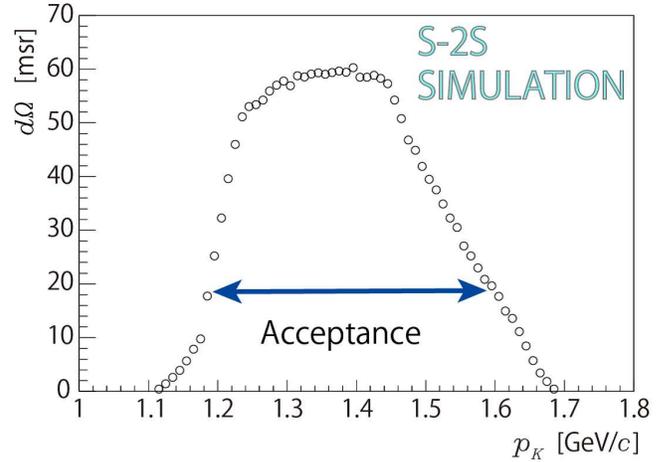}
    \caption{Momentum dependence of solid angle of S-2S, 
      obtained through a Monte Carlo simulation 
      as presented in Sec.~\ref{sec:epr_geant}. 
      An arrow represents a momentum coverage region with the S-2S
      solid angle of greater than 30$\%$ of that at 1.3~GeV/$c$.}
    \label{fig:solid_angle_s-2s}
  \end{center}
\end{figure}

A spectroscopic study of the $\Xi^{-}$ hypernucleus 
via the {\kk} reaction at an incident momentum of 1.8~GeV/$c$ (J-PARC E05)~\cite{cite:nagae}
is planned at the K1.8 beam line of the hadron experimental hall in J-PARC~\cite{cite:takahashi}.
The experiment is the first attempt to investigate the $\Xi^{-}$ 
hypernucleus, $^{12}_{\Xi}$Be, with an energy resolution of a few MeV (FWHM) by using a new 
magnetic spectrometer, Strangeness -2 Spectrometer (S-2S).
The S-2S consists of two quadrupole magnets (Q1, Q2) and one dipole magnet (D), 
as shown in Fig.~\ref{fig:s2s_setup}. 
The momentum acceptance of S-2S was estimated 
using the same simulation setting as that presented in Sec.~\ref{sec:epr_geant}, 
and a solid angle was obtained as a function of incident momentum, 
as shown in Fig.~\ref{fig:solid_angle_s-2s}. 
The solid angle for a particle at 1.3~GeV/$c$ is approximately 60~msr. 
A momentum coverage greater than 30$\%$ of the solid angle at 1.3~GeV/$c$ 
ranges from approximately 1.2 to 1.6~GeV/$c$.
The S-2S was designed to measure the $K^{+}$ momentum with a momentum 
resolution of $\Delta p/p \simeq 5.0\times 10^{-4}$ (FWHM) in the momentum 
range of 1.2 to 1.6~GeV/$c$.
The expected major background particles are protons and 
$\pi^{+}$s.
According to our Monte Carlo simulation based on the JAM code~\cite{cite:jam}, 
rates of background triggers from 
protons and $\pi^{+}$s approximately a thousand times higher than that of $K^{+}$s are expected 
in the S-2S acceptance. 
Therefore, background suppression at a trigger 
level (on-line) is essential to conduct the experiment 
in a given beam time while maintaining sustainable rates for our 
data acquisition system (DAQ). 

Background-particle suppression with {\cerenkov} detectors was
well established in past $\Lambda$ 
hypernuclear experiments in a similar momentum range~\cite{cite:sks1,cite:sks2,cite:gogami}.
In the J-PARC E05 experiment, background-particle suppression is 
also designed to be performed with a combination of {\cerenkov} detectors.
$\pi^{+}$s will be rejected with an existing aerogel (refractive index of 1.05) 
{\cerenkov} detector, which has been used in $\Lambda$ hypernuclear experiments 
with the Superconducting Kaon Spectrometer (SKS). 
On the other hand, a {\cerenkov} detector suppressing protons 
in the S-2S momentum and geometrical acceptance does not exist. 
In the present work, we developed a threshold-type {\cerenkov} detector 
with radiation medium of pure water (refractive index of 1.33) for 
proton suppression in S-2S, and the results of 
prototype tests are described here.



\section{Requirements for Water {\cerenkov} Detector}
Figure~\ref{fig:s2s_setup} shows a schematic drawing of the top view of S-2S.
The S-2S detector system consists of drift chambers for particle tracking, 
a time-of-flight (TOF) detector for trigger and off-line particle identification (PID),
and {\cerenkov} detectors for on-line PID.

\begin{figure*}[!htbp]
  \begin{center}
    \includegraphics[width=15cm]{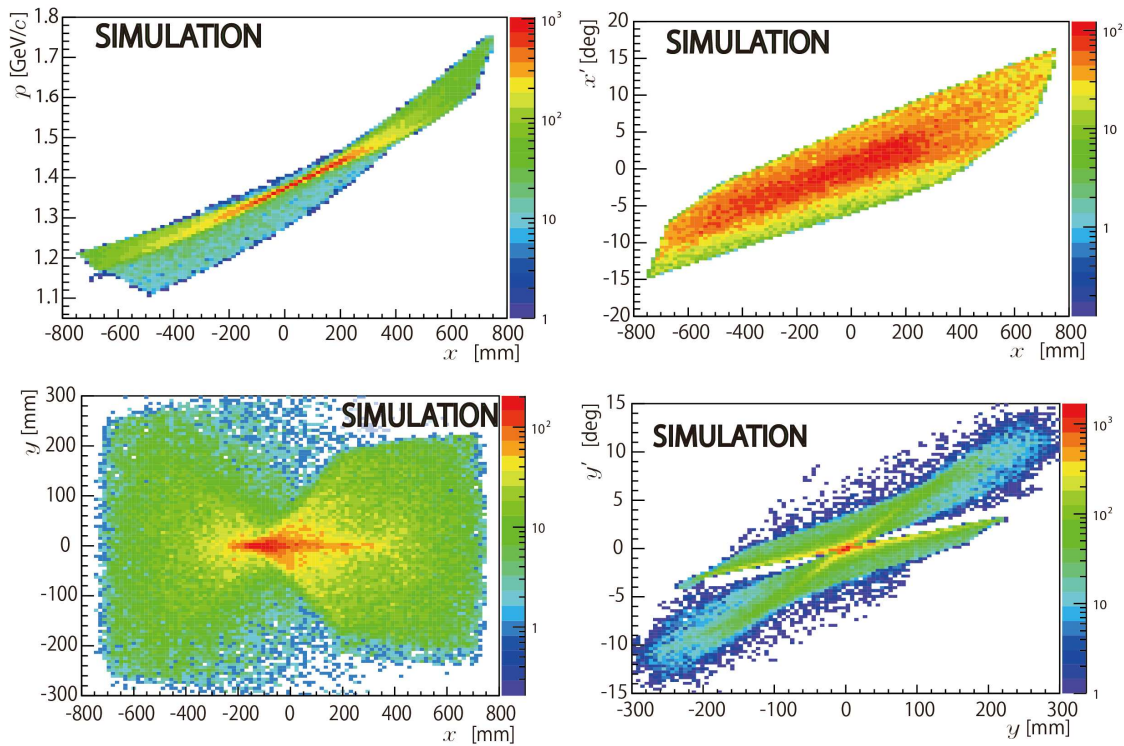}
    \caption{Particle distributions at a reference plane 
      3~cm upstream of the front surface of the water {\cerenkov} detector 
      in the S-2S Monte Carlo simulation (Sec.~\ref{sec:epr_geant}).}
    \label{fig:s2s_sim_distribution}
  \end{center}
\end{figure*}

A water {\cerenkov} detector 
with two layers to avoid dead spaces in each layer
(Fig.~\ref{fig:s2s_setup} and Fig.~\ref{fig:wc_12segments})
is planned for installation at the downstream end of the detector system.
The water {\cerenkov} detector is designed to be segmented into
six per layer. 
The segmented size was optimized for handling and maintenance, 
in addition to considerations of track multiplicity
and the number of photoelectrons in each segment.
Two photo-multiplier tubes (PMTs) are attached on each segment (Fig.~\ref{fig:wc_onebox}).
Analog pulses of the two PMTs are summed up, and 
the summed signal is input to a discriminator with a threshold avoiding protons 
to produce one of the trigger signals.
Figure~\ref{fig:s2s_sim_distribution} shows particle distributions 
(momentum vs. $x$, $x^{\prime} (\equiv \frac{dx}{dz})$ vs. $x$, $y$ vs. $x$, $y^{\prime} (\equiv \frac{dy}{dz})$ vs. $y$)
3~cm upstream of the water {\cerenkov} detector in S-2S obtained through a Monte Carlo simulation
(Sec.~\ref{sec:epr_geant}). 
We need to construct a water {\cerenkov} detector 
that accepts particles as shown in Fig.~\ref{fig:s2s_sim_distribution}.

\begin{figure}[!htbp]
  \begin{center}
    \includegraphics[width=8.6cm]{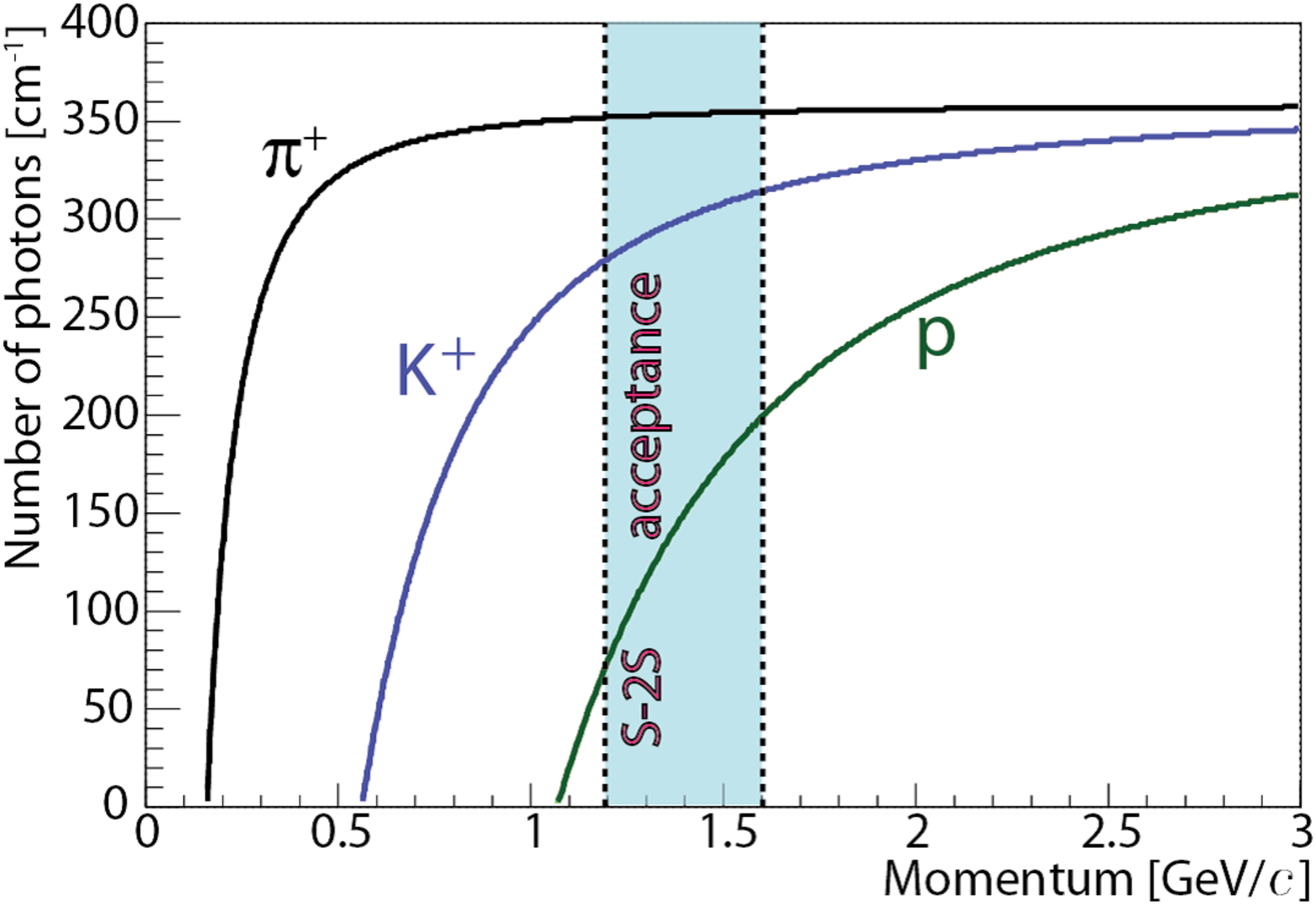}
    \caption{
      Photon yield of {\cerenkov} light per cm 
      in water ($n=1.33$)
      as a function of incident particle ($\pi^{+}$, $K^{+}$, proton) 
      momentum calculated using Eq.~(\ref{eq:wcnpe}).
    }
    \label{fig:n_photons}
  \end{center}
\end{figure}

S-2S measures momenta of particles with a range of 1.2--1.6~GeV/$c$. 
Figure~\ref{fig:n_photons} shows the photon yield of {\cerenkov} light per cm 
in water as a function of incident particle ($\pi^{+}$, $K^{+}$, $p$) 
momentum calculated as follows~\cite{cite:pdg}:
\begin{eqnarray}
  \frac{d^{2}N}{dxd\lambda } = \frac{2\pi \alpha z^{2}}{\lambda ^{2}}
  \Bigl(1-\frac{1}{\beta ^{2}n^{2}(\lambda)}\Bigr),
  \label{eq:wcnpe}
\end{eqnarray}
where $N$ and $\lambda$ are the number of photons and wavelength of {\cerenkov} light, respectively; 
$x$, $z$, and $\beta$ are the path length, charge, and velocity factor of the incident particle, respectively; 
and $n$($\lambda$) is the refractive index of the radiation medium. 
In Fig.~\ref{fig:n_photons}, the wavelength was integrated from 300~nm to 600~nm, which 
roughly corresponds to the light-sensitive region of a PMT, 
and $n$($\lambda$) is fixed at 1.33, which is the nominal refractive index of pure water.
All three particles yield {\cerenkov} light in water in the 
S-2S momentum acceptance, but can be identified by a difference of photon yield.
Protons are planned to be suppressed at a trigger level (on-line) by 
cutting the photon yield of the water {\cerenkov} detector. 

\begin{figure}[!htbp]
  \begin{center}
    \includegraphics[width=8.6cm]{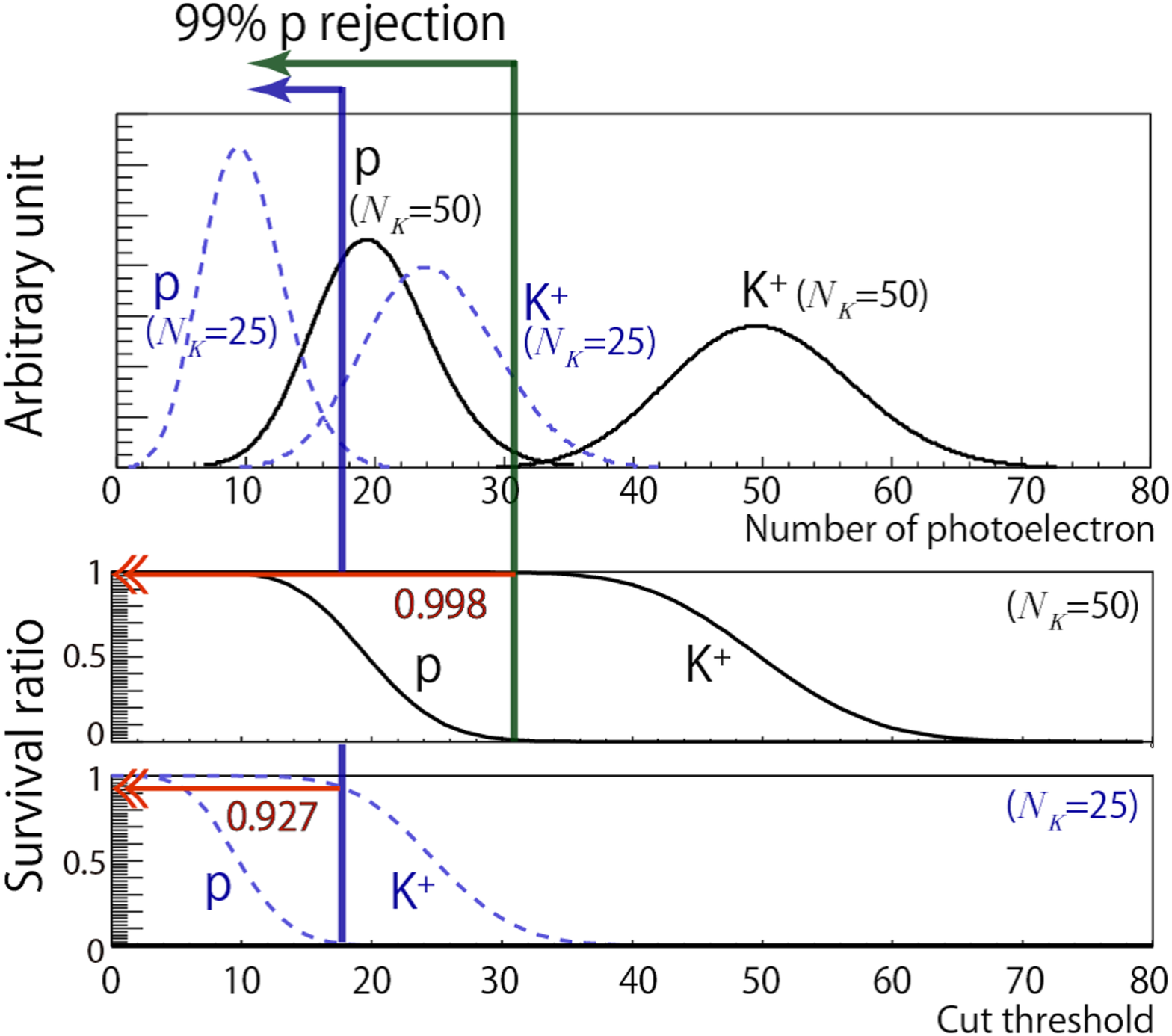}
    \caption{Top figure shows the expected distribution 
      (Poisson distribution) of the number of photoelectrons for 
      protons and $K^{+}$s with the momentum of 1.3~GeV/$c$, assuming 
      that the mean values for 1.3~GeV/$c$ $K^{+}$ 
      are 25 (dashed line, $N_{K}=25$) and 50 (solid line, $N_{K}=50$).
      Bottom two figures show survival ratios 
      of protons and $K^{+}$s as a function of threshold.
      If one chooses a threshold rejecting 99$\%$ of protons, 
      the survival ratios of $K^{+}$ for $N_{K}=25$ and $N_{K}=50$ are, 
      respectively, 92.7$\%$ and 99.8$\%$.}
    \label{fig:wc_fom}
  \end{center}
\end{figure}

{\cerenkov} light is converted to electrons (photoelectrons) 
and multiplied by a PMT to be detected as signals. 
Photon yield, which is measured as 
the number of photoelectrons (NPE), depends on 
the container shape of the radiation medium, 
window material used between the radiation 
medium and PMT, reflection material at the inside of the container, 
and PMT performance {\it etc.}
The top panel of Fig.~\ref{fig:wc_fom} shows the expected NPE distributions
assuming simple Poisson distributions for 
proton and $K^{+}$ with a momentum of 1.3~GeV/$c$, 
in two cases when the mean values for 1.3~GeV/$c$ $K^{+}$ 
are 25 (dashed line, $N_{K}=25$) and 50 (solid line, $N_{K}=50$).
In the bottom two panels of Fig.~\ref{fig:wc_fom}, survival ratios 
of protons and $K^{+}$s as a function of threshold
are shown. If one chooses a threshold rejecting 99$\%$ of protons, 
the survival ratios of $K^{+}$s for $N_{K}=25$ and $N_{K}=50$ are 
92.7$\%$ and 99.8$\%$, respectively.
As described above, a higher $K^{+}$ survival ratio can be achieved 
with a detector having a larger NPE detection power, 
when the proton rejection efficiency is fixed. 

In the J-PARC E05 experiment, 
we plan to use a $10^{6}$~counts/spill 
$K^{-}$ beam\footnote{Two seconds of flat top is expected at the moment.} 
at the K1.8 beam line in J-PARC. 
The counting rates of protons, $K^{+}$s, and $\pi^{+}$s 
from a 3.0~g/cm$^{2}$ $^{12}$C target are estimated to be 150, 1, and 
20~[counts/$10^{6}$ $K^{-}$ beams], respectively, 
according to a Monte Carlo simulation (Geant4~\cite{cite:geant4}). 
Background particles, which do not originate from the 
target, such as decay particles
from  $K^{-}$ beams, are also simulated by the 
Monte Carlo method, and the counting rate is 
estimated to be a few tens~[counts/$10^{6}$ $K^{-}$ beams].
We have confirmed that the rate estimation is not far
from the reality in S-2S, because the Monte Carlo simulation is able to reproduce 
particle rates (proton, $\pi^{+}$, and $\gamma$) 
in a test experiment by using the SKS with $K^{-}$ beams
at J-PARC in 2015. 
A safety factor of five was multiplied to the estimated 
background counting rate
($\simeq 200 \times 5 = 1000$ [counts / $10^{6}$ $K^{-}$ beams]),
and this value was used for the S-2S detector design.
A DAQ trigger rate would be desirable to maintain the rate below a few hundred counts/spill
in order to obtain data with sufficient efficiency.
The existing aerogel {\cerenkov} detector is able 
to reject $\geq 99.7\%$ of $\pi^{+}$s
with a $K^{+}$ survival 
rate of approximately $93\%$ at the trigger level~\cite{cite:ota,cite:ichikawa}.
Therefore, a
proton-rejection efficiency of $\geq 90\%$, 
keeping a $K^{+}$ survival rate of $\geq 95\%$, is the goal 
for the water {\cerenkov} detector at the trigger level.


It is noted that a PMT gain is deteriorated 
by a magnetic field. 
The water {\cerenkov} detector is planned to be installed 
a few meters away from the S-2S dipole magnet. 
We measured the magnetic field around positions 
where PMTs of the water {\cerenkov} detector
will be, and found that 
there is about $B_{y}=5$~G
at maximum\footnote{$B_{y}$: magnetic field parallel to the axis of PMT.}.
In the study of \cite{cite:gogami}, 
the gain of a PMT (Hamamatsu H7195),
which is the similar PMT type to ours (Hamamatsu H11284-100UV), 
is reduced by $60\%$ in the case that 
the magnetic field of $B_{y}=5$~G is yielded on the PMT.
Therefore, the magnetic field on the 
PMTs should be suppressed in order to avoid 
a deterioration of the proton-rejection power which 
is caused by PMT-gain reductions.
For the water {\cerenkov} detector in S-2S, 
we plan to adopt the bucking coil method 
which was proven to work to recover the PMT gain 
against the similar strength of magnetic field
($B_{y}\simeq 4$--6~G)
in an experiment of $\Lambda$ hypernuclear spectroscopy
with the {\eek} reaction~\cite{cite:gogami}.

\section{Prototype Water {\cerenkov} Detector}
\label{sec:old_pwc}
A prototype water {\cerenkov} detector was constructed and 
tested by irradiating positron beams at 
the Research Center for Electron Photon Science, Tohoku University (ELPH) 
to study the basic performance of the prototype.
In this section, the prototype design and results of the beam test are described.

\subsection{Design of the Prototype Water {\cerenkov} Detector}
\label{sec:prototype_design0}
A water {\cerenkov} detector was used in experiments 
of $\Lambda$ hypernuclear spectroscopy with the {\eek} reaction 
at Jefferson Lab (JLab)~\cite{cite:gogami}.
There are accumulated studies about the water {\cerenkov} 
detector in the {\eek} experiment at JLab~\cite{cite:wc_theses},
and our design and choices of materials for a prototype
were inspired by them~\cite{cite:wc_theses2}.

\begin{figure}[!htbp]
  \begin{center}
    \includegraphics[width=8.6cm]{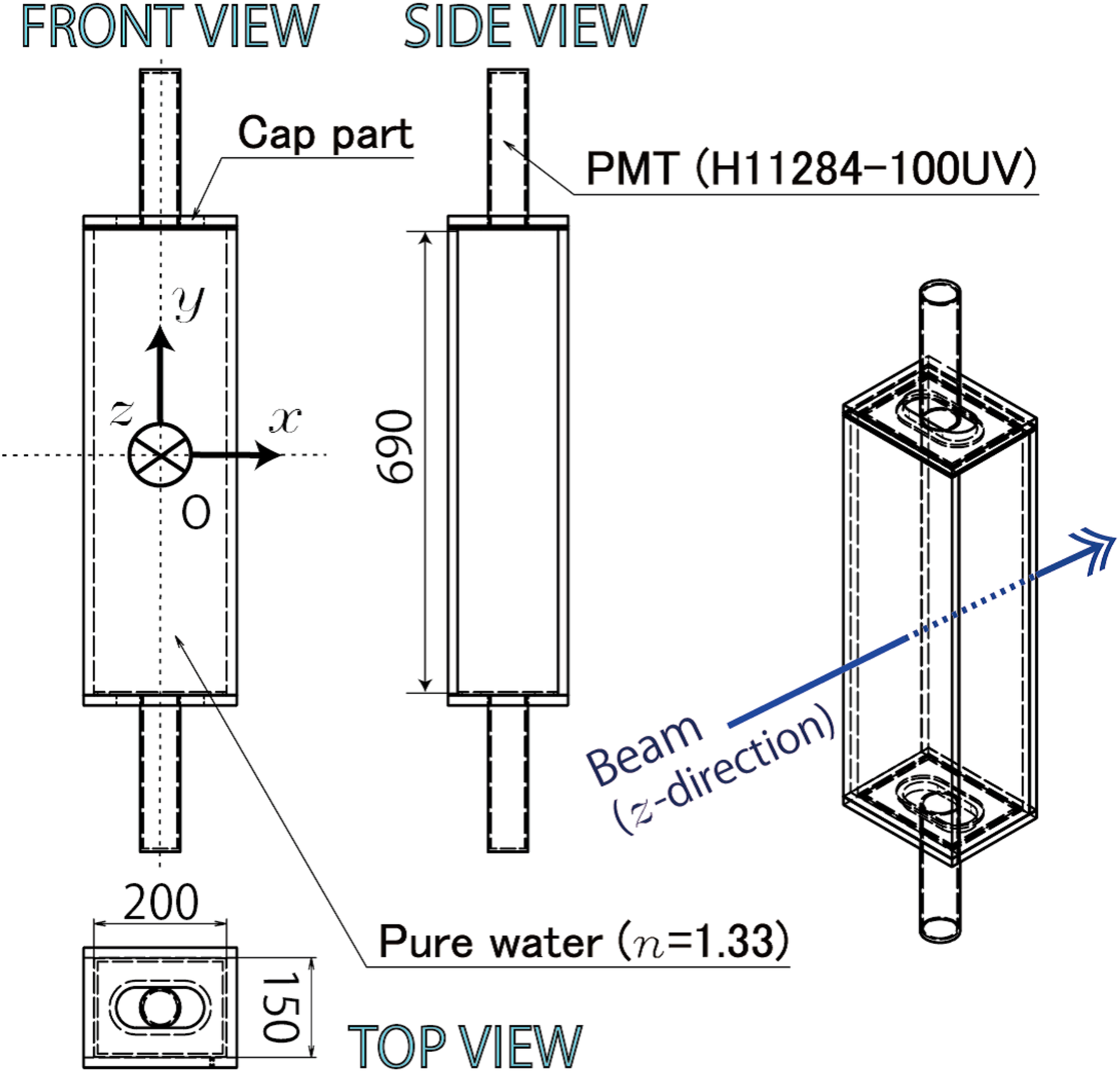}
    \caption{Drawing of the prototype water {\cerenkov} detector.
      $x$, $y$, and $z$ coordinates are defined in the drawing. All 
      dimensions are in mm.}
    \label{fig:wc_onebox}
  \end{center}
\end{figure}

The prototype water {\cerenkov} detector consists of 
a white acrylic container filled with pure water 
(purified water Clean $\&$ Clean\footnote{KOGA Chemical Mfg Co., Ltd., http://www.kykk.jp/}, 
electrical conductivity of $\leq 0.1$~mS/m at 25$^{\circ}$C)
with a transparent acrylic window
attached to a PMT photocathode at the top/bottom side.
The container was made of 15-mm-thick white acrylic boards 
bonded by polymerization bonding, except for a cap part. 
The cap part was attached to the main container 
by winding with polypropylene bands (PP bands).
A diffuse reflection material, 
Tyvek sheet (Tyvek 1060B\footnote{DuPont, http://www.dupont.com/}),
which is widely used as a building material,
was attached on the inside of the container by a support structure. 
PMTs (Hamamatsu 
H11284-100UV~\cite{cite:hamamatsu})
were attached 
on the top and bottom of the container by using silicon optical coupling grease
(BC-630\footnote{Saint-Gobain Crystals, https://www.saint-gobain.com/fr}, $n=1.465$). 
Between the radiation medium and the PMT,
5~mm windows made of transparent acrylics 
(Acrylite$\#$001\footnote{Mitsubishi Rayon company, http://www.mrc.co.jp/acrylite/})
were inserted.
After the above materials were assembled, 
we winded the black sheets (thickness of 0.1~mm), 
which is made of polyvinyl chloride (PVC),  
around the container more than twice for light shielding.
Figure~\ref{fig:wc_onebox} shows a 
drawing of the prototype water {\cerenkov} detector.
The effective volume was $200^{W} \times 690^{H} \times 150^{T}$~mm$^{3}$.

\subsection{Experimental Setup}
\label{sec:elphexp}
A test experiment was performed using positron beams 
in the positron/electron beam line at 
ELPH~\cite{cite:ishikawa1,cite:ishikawa2}
in order to investigate the basic 
performance factors of the prototype water {\cerenkov} detector
such as position and angular dependencies of NPE. 

\begin{figure}[!htbp]
  \begin{center}
    \includegraphics[width=8.6cm]{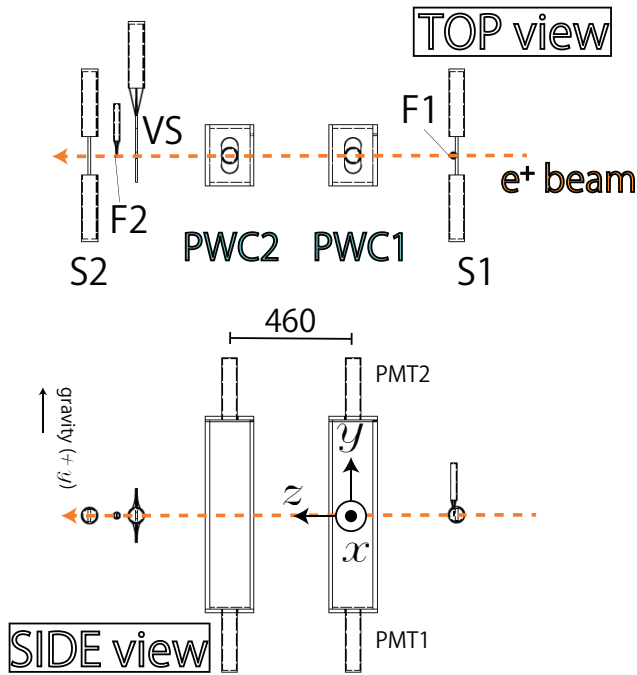}
    \caption{Schematic drawing of the experimental setup at ELPH.
      Positron beams with $\beta=1$ were incident on two sets of prototype water 
      {\cerenkov} detectors (PWC). The distance in the figure is in mm.}
    \label{fig:wc_elph_setup}
  \end{center}
\end{figure}

Figure~\ref{fig:wc_elph_setup} shows a schematic drawing 
of the experimental setup. 
Positron beams with $\beta=1$ ($\beta > 0.99999$) impinged on 
two sets of prototype water {\cerenkov} detectors (PWC1, PWC2).
Two plastic scintillation detectors (S1, F1) 
before PWCs and three plastic 
scintillation detectors (VS, F2, S2) after PWCs were installed. 
VS had a $\phi$25~mm hole at the center
so as to reject background particles in the off-line analysis.
S1, S2, and F1 were used for both data-taking trigger
and off-line analysis.
Detector sizes and distances with respect to S1 
are summarized in Table~\ref{tab:detector_size}.
\begin{table*}[!htbp]
  \begin{center}
    \caption{Size and distance from S1 in the beam direction of 
      each detector used in the test experiment at ELPH.}
    \label{tab:detector_size}
    \begin{tabular}{|c|c|c|c|}
      \hline \hline
      Detector &  Distance  & Size [mm] & Remarks \\ 
               &   [mm]     &           &         \\ \hline \hline
      S1 & 0 & $144^{W} \times 44^{H} \times 10^{T}$ &  Plastic scintillation detector \\
      & & & (used for trigger) \\ \hline
      F1 & 6 & $9.3^{W} \times 9.3^{H} \times 4^{T}$ & Plastic scintillation detector \\
      & & & (used for trigger) \\ \hline
      PWC1 & 382 & $200^{W} \times 690^{H} \times 150^{T}$ & Water {\cerenkov} detector\\ \hline 
      PWC2 & 842& $200^{W} \times 690^{H} \times 150^{T}$ & Water {\cerenkov} detector\\ \hline
      VS & 1190&  $200^{W} \times 200^{H} \times 5^{T}$& Plastic scintillation detector\\
      & & & with a $\phi$25 hole at the center \\ 
      & & & (used for off-line analysis) \\ \hline
      F2 & 1266& $9.8^{W} \times 9.5^{H} \times 4^{T}$ & Plastic scintillation detector \\
      & & & (used for off-line analysis) \\ \hline
      S2 & 1369& $144^{W} \times 44^{H} \times 10^{T}$ & Plastic scintillation detector \\ 
      & & & (used for trigger) \\ \hline \hline
    \end{tabular}
  \end{center}
\end{table*}

PWCs could be displaced and tilted using 
a movable frame in order to change incident positions and angles
of the beams on PWCs. $x$- and $y$-coordinates 
are defined in Fig.~\ref{fig:wc_elph_setup}, and 
$x^{\prime} (\equiv \frac{dx}{dz})$ and $y^{\prime} (\equiv \frac{dy}{dz})$ are the angles of 
incident beams on PWC.

\subsection{Analysis}
\subsubsection{NPE calibration}
\label{sec:led_calib}
The number of photoelectrons was obtained from the analog-to-digital converter (ADC) 
spectrum of each PMT of the water {\cerenkov} detector.
The ADC was calibrated with light from a light-emitting diode (LED) 
(Para Light LED 3.0~mm L-314LBD\footnote{Para Light Electronics company, http://www.para.com.tw/}),
which was installed in the container of the
water {\cerenkov} detector. 
The LED light intensity was adjusted to generate a few NPE signals
on each PMT by controlling the height and width of 
input rectangular pulses generated by an arbitrary/function generator.
\begin{figure}[!htbp]
  \begin{center}
    \includegraphics[width=8.6cm]{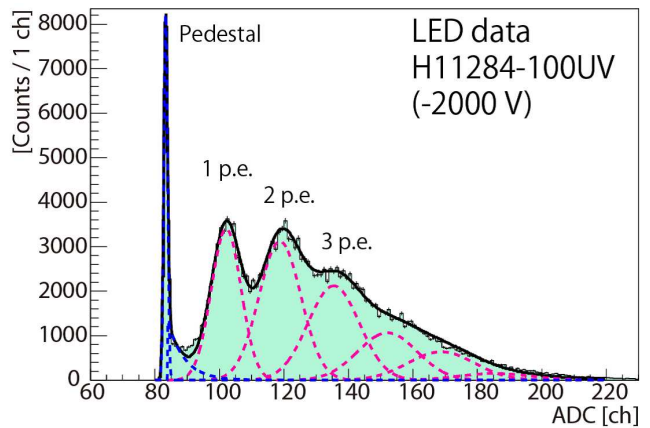}
    \caption{ADC histogram for a PMT, H11284-100UV, using LED.
      The ADC histogram was fitted with  
      the convolution of a Gaussian function for the pedestal, 
      Poisson functions for photoelectron peaks, and an exponential function for 
      residual background events between the pedestal and a single photoelectron 
      peak (1 p.e.)~\cite{cite:wcfit}.}
    \label{fig:wc1LED}
  \end{center}
\end{figure}
Fig.~\ref{fig:wc1LED}
shows an ADC histogram of H11284-100UV obtained using the LED light.
The ADC histogram was fitted with a 
convolution of a Gaussian function for the pedestal, 
Poisson functions for photoelectron peaks, and an exponential function for 
residual background events between the pedestal and a single photoelectron 
peak (1 p.e.)~\cite{cite:wcfit}. 
The ADC channel for a single photoelectron ($A_{{\rm s.p.e}}$)
was then obtained from the fitting, and it 
was used for conversion from an ADC histogram to an NPE histogram as follows:
\begin{eqnarray}
  {\rm NPE} = \frac{{\rm ADC}-{\rm Pedestal}}{A_{{\rm s.p.e}}}.
\end{eqnarray}

\subsubsection{Mean NPE Derivation}
\begin{figure}[!htbp]
  \begin{center}
    \includegraphics[width=8.6cm]{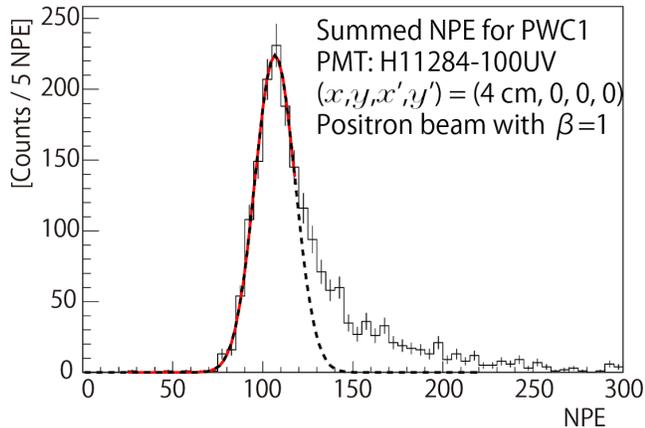}
    \caption{Typical NPE histogram with 
      a fitting result after the event selection 
      of Eq.~(\ref{eq:elph_selection})
      and PWC TDC selection were applied.
      The mean NPE value obtained by the fitting was used as a result.}
    \label{fig:npe_elph}
  \end{center}
\end{figure}

In order to obtain a mean NPE value (MNPE), the converted NPE histogram was 
fitted with a Gaussian function.  An event selection was applied in the 
NPE analysis as follows: 
\begin{eqnarray}
  \label{eq:elph_selection}
  S1 \otimes S2 \otimes F1 \otimes F2 \otimes \overline{VS},
\end{eqnarray}
where S1, S2, F1, F2, and VS correspond to ADC and TDC 
selections for each detector.
In addition, TDC\footnote{Time to Digital Converter}
selections of PWCs were also applied 
to select events in an appropriate timing window. 
Fig.~\ref{fig:npe_elph} shows a typical NPE histogram 
with a fitting result after the above event selections were applied.

\subsection{Results}
\subsubsection{Mean NPE at Center position}
\label{sec:npe_elph_center}
The mean NPEs obtained for PWC1 and PWC2 
at $(x,y,x^{\prime},y^{\prime})=(0,0,0,0)$ 
are summarized in Table~\ref{tab:elph_result_y0}.
\begin{table}[!htbp]
  \begin{center}
    \caption{Mean NPEs (MNPE, MsNPE) for PWC1 and PWC2 when 
      positron beams impinged at $(x,y,x^{\prime},y^{\prime})=(0,0,0,0)$.
      Errors in the results are statistical.}
    \label{tab:elph_result_y0}
    \begin{tabular}{|c|l|cc|}
      \hline \hline
      \multicolumn{2}{|c|}{ } & PWC1 & PWC2 \\ \hline \hline
      MNPE  &PMT1 (top)     & $64.8 \pm 0.3$ & $65.2 \pm 1.0$\\
            &PMT2 (bottom)  & $41.2 \pm 0.2$ & $65.6 \pm1.9$ \\  \hline
      MsNPE &PMT1+PMT2 (sum)& $105.7 \pm 0.4$ & $135.0\pm0.8$\\ \hline \hline
    \end{tabular}
  \end{center}
\end{table}
The MNPE detected in each PMT was approximately 65, except for the bottom PMT of PWC1 (PWC1-2). 
The smaller NPE of PWC1-2 could be possibly caused by an air gap 
between the PMT and acrylic window. 
As described in Sec.~\ref{sec:grease_eff}, an air gap 
causes an NPE reduction of approximately $35\%$. 
If the NPE of PWC1-2 is corrected by a factor of $f_{{\rm gap}}=\frac{1}{(1.00-0.35)}$, 
the NPE becomes NPE$^{{\rm PMT}1-2}\times{f_{{\rm gap}}}=63.4 \pm 0.3$,
which is then consistent with the others within the 
standard deviation of PMT individual performance difference 
($\sigma_{{\rm PMT}}=6\%$, Sec.~\ref{sec:pmt_ind_diff}). 
The air gap was considered to be caused by loose PMT attachment to the frame.

\subsubsection{Incident Angle ($y^{\prime}$) Dependence}
Mean summed NPEs\footnote{Summed NPE; a mean NPE obtained by fitting to a summed NPE spectrum
(NPE sum of top and bottom PMTs).} (MsNPE) of PWC1 were obtained as follows:
\begin{eqnarray}
  {\rm MsNPE}^{y^{\prime}=0^{\circ}}_{{\rm center}} &=& 135.0 \pm 0.8 \hspace{0.1cm},
  \label{eq:yp_elph}\\ 
  {\rm MsNPE}^{y^{\prime}=8^{\circ}}_{{\rm center}} &=& 84.3\pm0.5,  \label{eq:yp_elph_}
\end{eqnarray}
at $(x,y,x^{\prime},y^{\prime})=(0,0,0,0)$ and $(0,0,0,8^{\circ})$, respectively.
There was a large reduction in ${\rm MsNPE}^{y^{\prime}=8^{\circ}}_{{\rm center}}$
compared to ${\rm MsNPE}^{y^{\prime}=0^{\circ}}_{{\rm center}}$.
The NPE reduction would be caused by the 
detachment of PMTs from acrylic windows when the PWC was tilted to observe 
the angular dependence as well as the situation of 
PWC1-2 in Sec.~\ref{sec:npe_elph_center}. 
If the ${\rm MsNPE}^{y^{\prime}=8^{\circ}}_{{\rm center}}$ is corrected 
by $f_{{\rm gap}}$: 
\begin{eqnarray}
  {\rm MsNPE}^{y^{\prime}=8^{\circ}}_{{\rm center}} \times f_{{\rm gap}}
  = 130.0 \pm 0.8 \hspace{0.1cm}. \label{eq:yp_cor_elph}
\end{eqnarray}
The MsNPE variation was obtained to be approximately $4\%$ in the range 
from $y'=0$ to $8$~deg 
by comparing the two values in Eq.~(\ref{eq:yp_elph}) and Eq.~(\ref{eq:yp_cor_elph}). 


\subsubsection{Vertical Position ($y$) Dependence}
\begin{figure}[!htbp]
  \begin{center}
    \includegraphics[width=8.6cm]{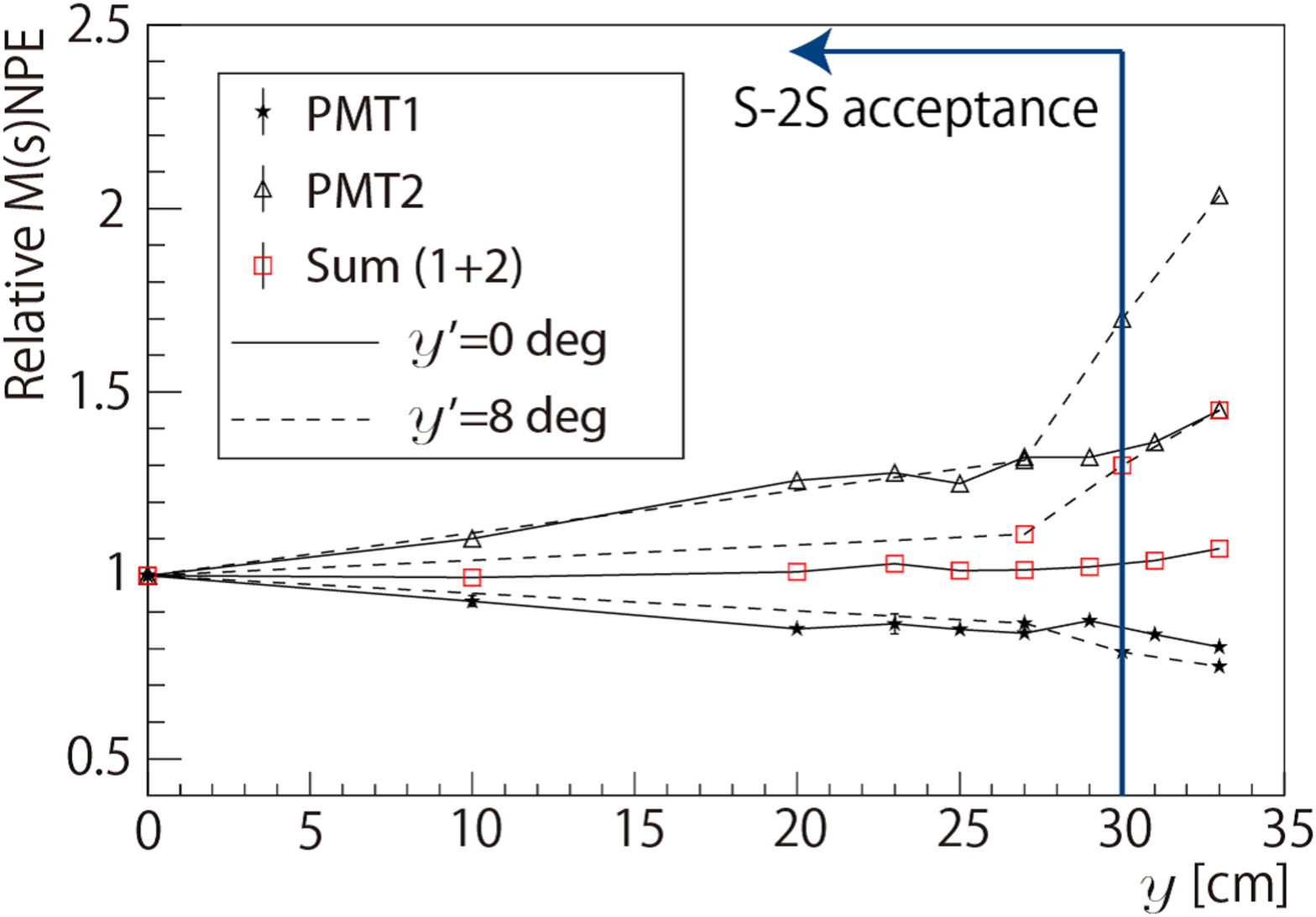}
    \caption{Relative MNPEs and MsNPEs as functions of $y$-position 
      at $y^{\prime}=0$ and 8~deg.}
    \label{fig:ydep_ang0}
  \end{center}
\end{figure}

Figure~\ref{fig:ydep_ang0} shows 
the $y$-position dependence of M(s)NPE at $(x,x^{\prime})=(0,0)$ 
for $y^{\prime}=0$ and 8~deg.
The M(s)NPE values were normalized so as to make 
the M(s)NPEs at $(x,x^{\prime},y)=(0,0,0)$ unity. 

\begin{figure*}[!htbp]
  \begin{center}
    \includegraphics[width=13cm]{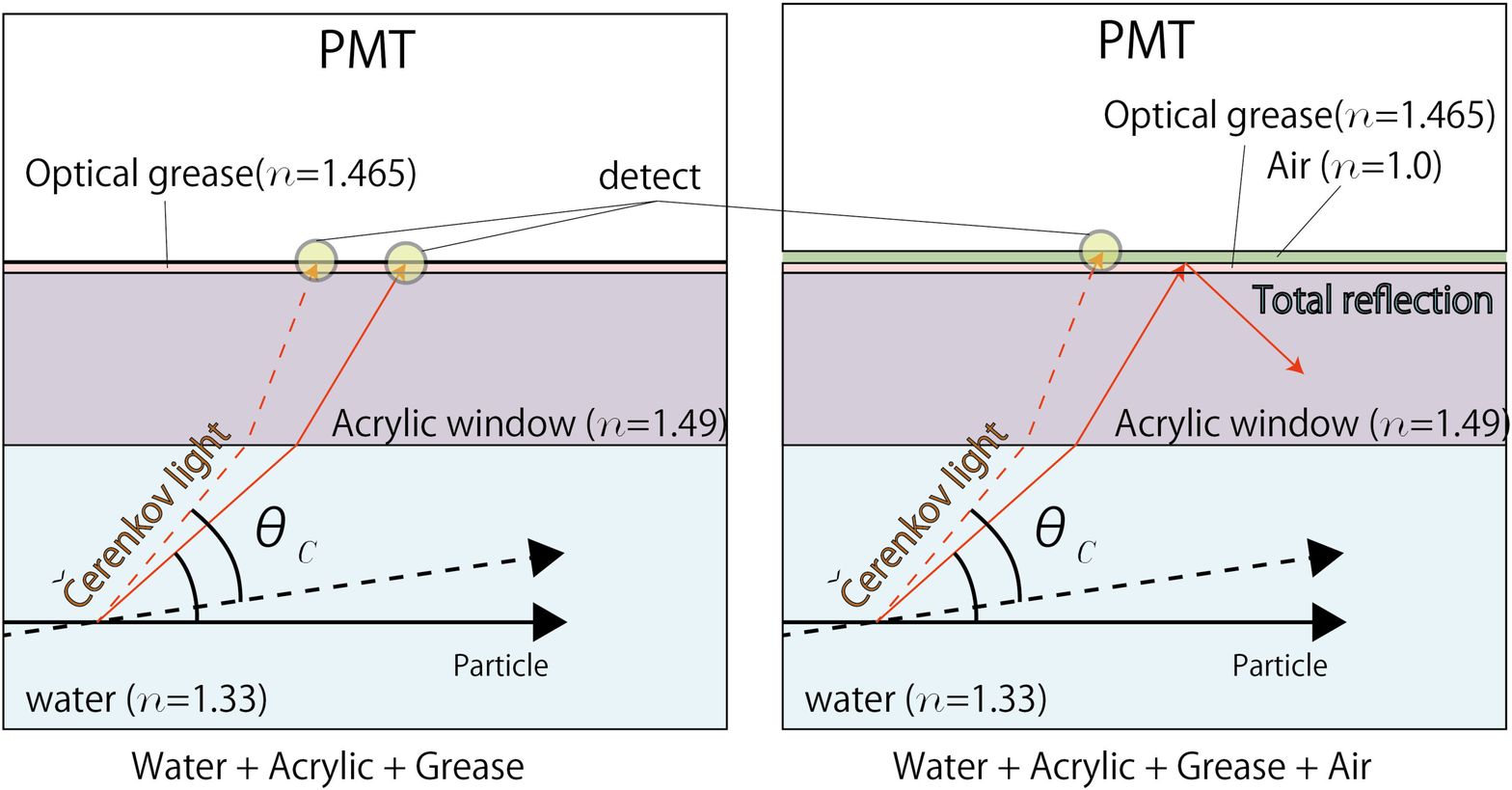}
    \caption{Schematic diagrams for direct {\cerenkov} light detection by PMT.
      Total reflection occurs for $y^{\prime}=0$~deg 
      before the PMT photocathode
      if there is an air gap between the optical coupling grease and PMT, as 
      described in the right figure.}
    \label{fig:Cerenkov_angle}
  \end{center}
\end{figure*}

The MNPE of PMT2 for $y^{\prime}=8$~deg steeply increases
at approximately $y=28$~cm. 
This is considered to be caused by 
direct {\cerenkov} light detection in the PMT. 
The {\cerenkov} radiation angle relative to an
incident particle direction is calculated as follows~\cite{cite:pdg}: 
\begin{eqnarray}
  \theta_{C} = \arctan ( \sqrt{\beta^{2}n^{2}-1} ).
\end{eqnarray}
Thus, the radiation angle of a $\beta=1$ particle in water ($n=1.33$) 
is obtained as follows:
\begin{eqnarray}
  \theta^{\beta=1}_{C} = 41.2^{\circ}.
  \label{eq:theta_cerenkov}
\end{eqnarray}
Taking into account the radiation angle,
the geometry of the prototype water {\cerenkov} detector 
and the refraction between water and the acrylic window ($n=1.49$),
the direct {\cerenkov} light arrives
at the PMT photocathode at $y=24$~cm and $y=27$~cm for 
$y^{\prime}=0$ and 8~deg, respectively. 
In the results, however, 
the NPE increase starts at approximately $y=28$~cm for $y^{\prime}=8$~deg.  
The difference might be due to the air gap
between the optical coupling grease and PMT.
Moreover, there is no such increase for $y^{\prime}=0$~deg.
This could also be explained by the air gap. 
{\cerenkov} light directly reaches 
the PMT photocathode if there is no air gap, 
as shown in the left schematic diagram of Fig.~\ref{fig:Cerenkov_angle}.
On the other hand, 
{\cerenkov} light that satisfies the following condition
does not reach the PMT photocathode directly owing to total reflection 
at a boundary between the air and optical coupling grease:
\begin{eqnarray}
  \theta_{{\rm in}} \geq 90^{\circ} - \arcsin \Bigl( \frac{n_{2}}{n_{1}} \Bigr) = 41.2^{\circ} ,
  \label{eq:theta_total-reflection}
\end{eqnarray}
where $\theta_{{\rm in}}$ ($0^{\circ} \leq \theta_{{\rm in}}<90^{\circ}$) is the angle of 
light with respect to the $z$-axis on the $yz$-plane and
$n_{1}$ and $n_{2}$ are, 
respectively, the refractive indexes of water ($n=1.33$) and air ($n=1.00$).
In the calculation, it is assumed that the thickness of 
the optical grease is zero.
Using Eq.~(\ref{eq:theta_cerenkov}), 
$\theta_{{\rm in}}$ for $y^{\prime}=0$~deg and $y^{\prime}=8$~deg are 
obtained as follows: 
\begin{eqnarray}
  \theta^{y^{\prime}=0^{\circ}}_{{\rm in}} 
  = \theta^{\beta=1}_{C} - 0^{\circ} = 41.2^{\circ}, 
  \label{eq:zero_reflect}\\
  \theta^{y^{\prime}=8^{\circ}}_{{\rm in}} 
  = \theta^{\beta=1}_{C} - 8^{\circ} = 33.2^{\circ}. 
  \label{eq:eight_reflect}
\end{eqnarray}
According to Eq.~(\ref{eq:theta_total-reflection}), 
Eq.~(\ref{eq:zero_reflect}), and Eq.~(\ref{eq:eight_reflect}), 
total reflection occurs for $y^{\prime}=0$~deg, 
but does not for $y^{\prime}=8$~deg, 
as shown in the right schematic diagram of Fig.~\ref{fig:Cerenkov_angle}, which 
is consistent with what we observed in Fig.~\ref{fig:ydep_ang0}.

\begin{figure}[!htbp]
  \begin{center}
    \includegraphics[width=8.6cm]{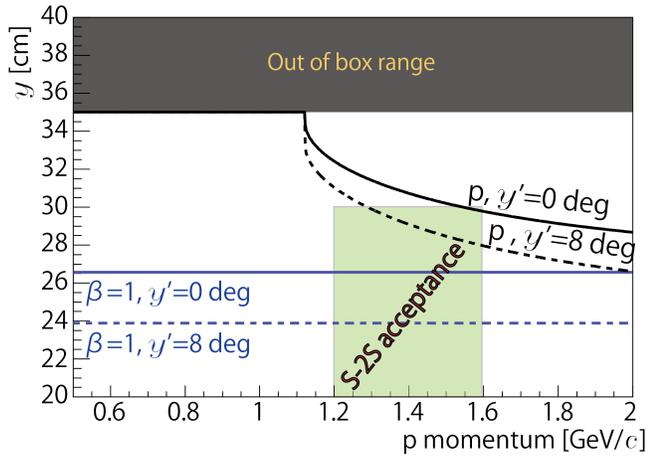}
    \caption{Calculated $y$ thresholds ($y^{{\rm thr}}$) to detect direct {\cerenkov} light on PMT in
      the prototype water {\cerenkov} detector for incident particles at $y^{\prime}=0$ and 8~deg.
      The $y^{{\rm thr}}$ for particles with $\beta=1$ is also shown as a reference.}
    \label{fig:Cerenkov_angle_threshold}
  \end{center}
\end{figure}

In the S-2S experiment, the proton-rejection efficiency 
would be reduced if direct {\cerenkov} light detection occurs while
keeping the same $K^{+}$ detection efficiency.
Figure~\ref{fig:Cerenkov_angle_threshold} shows 
a $y$ threshold ($y^{{\rm thr}}$) for {\cerenkov} light directly arriving at 
a PMT photocathode as a function of proton momentum for $y^{\prime}=0$ and 8~deg. 
The $y^{{\rm thr}}$ for particles with $\beta=1$ is also shown as a reference.
The S-2S acceptance is shown as a colored box, and 
the thresholds of protons 
are barely in the acceptance.
However, the ratio of the number of events that would cause 
direct {\cerenkov} light detection to those in the whole S-2S acceptance 
is estimated to be $<0.1\%$ according to 
a Monte Carlo simulation.
Therefore, the effect of reduction in proton rejection efficiency due 
to direct {\cerenkov} light detection is considered to be 
negligibly small.

\subsubsection{$x,y$-Position Dependence}
\begin{figure}[!htbp]
  \begin{center}
    \includegraphics[width=8.6cm]{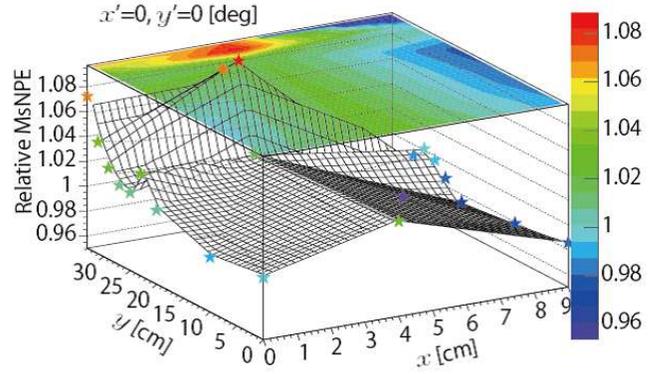}
    \caption{Dependence of the normalized MsNPE on $x$ and $y$ 
      detected by the prototype water {\cerenkov} detector 
      at $(x^{\prime},y^{\prime})=(0,0)$.}
    \label{fig:xydep_2d}
  \end{center}
\end{figure}

Figure~\ref{fig:xydep_2d} shows an MsNPE histogram 
in two-dimensions for $(x^{\prime},y^{\prime})=(0,0)$.
The MsNPE at $(x,y)=(0,0)$ is normalized to unity in the figure.
It was found that the MsNPE is higher for larger $|y|$ and lower for larger $|x|$. 
The incident-beam-position dependence of the MsNPE was measured to be $< \pm10\%$.

\subsubsection{Incident Angle ($x^{\prime}$) Dependence}
The horizontal beam angle, 
$x^{\prime}$, was also varied in the beam test ($0 \leq x^{\prime}\leq 8$~deg),
but no dependence on $x^{\prime}$ was observed within the error.

\subsubsection{Summary of the Beam Test}
In summary of the $\beta=1$ positron beam test, 
the prototype water {\cerenkov} detector 
was able to achieve a summed NPE of $\geq130$ for a $\beta=1$ particle.
An MsNPE variation of $<\pm10\%$ due to 
the beam position and angular dependencies is confirmed
for almost all particle tracks to be measured in the S-2S acceptance.

\section{Selection of {\cerenkov} Light Window Material}
\label{sec:window_material_opt}
Based on the measured NPE mean value for the previous 
prototype, the proton-rejection efficiency was estimated 
by using a Monte Carlo simulation. 
The Monte Carlo simulation showed that the 
proton rejection efficiency at the higher momentum region in S-2S 
would not be sufficient.
We attempted to improve the NPE detection power
by optimizing the choice of {\cerenkov} light window material. 

\subsection{Motivation}
\begin{figure}[!htbp]
  \begin{center}
    \includegraphics[width=8.6cm]{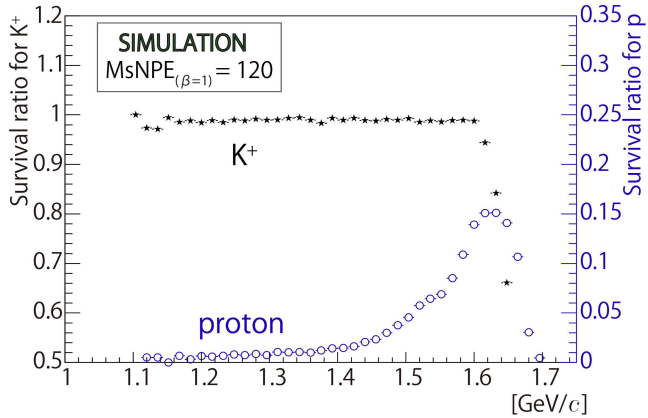}
    \caption{Estimated survival ratios for protons and $K^{+}$s in S-2S
       as functions of momentum at the particle generation point,
       assuming ${\rm MsNPE}=120$ for $\beta=1$ particles.}
    \label{fig:suvratio_npe120}
  \end{center}
\end{figure}

We achieved ${\rm MsNPE} \geq 130$ in the prototype water {\cerenkov} detector
(PWC) for a $\beta=1$ particle, as shown in the previous section.
A Monte Carlo simulation was performed to estimate the proton-rejection efficiency
with the same simulation framework as that described in Sec.~\ref{sec:epr_geant}, 
taking into account the obtained MsNPE and its dependence on incident-particle angles and positions. 
Figure~\ref{fig:suvratio_npe120} shows estimated survival ratios for 
protons and $K^{+}$s as a function of particle momentum in S-2S
assuming that  ${\rm MsNPE}=120$\footnote{The assumed NPE value 
is smaller by approximately 10$\%$ than that obtained in the positron beam test.} 
for $\beta=1$ particles.
The proton-rejection efficiency at approximately 1.6~GeV/$c$
is less than our goal of $90\%$ when maintaining the $K^{+}$ survival ratio 
of $\geq 95\%$. 
Thus, further development to improve NPE was needed.

\subsection{Effect of Window Material on NPE}
\begin{figure}[!htbp]
  \begin{center}
    \includegraphics[width=8.6cm]{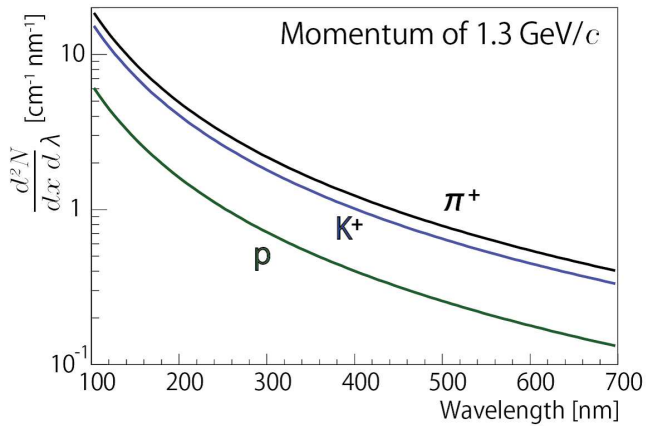}
    \caption{Number of photons of {\cerenkov} light 
      per unit path length of incident particle per 
      wavelength as a function the wavelength 
      for proton, $K^{+}$, and $\pi^{+}$ with the momentum of 1.3~GeV/$c$.}
    \label{fig:np_wavelength}
  \end{center}
\end{figure}

Figure~\ref{fig:np_wavelength} shows the calculated number of photons 
of {\cerenkov} light per unit length of incident particle per 
wavelength as a function of 
the wavelength for proton, $K^{+}$, and $\pi^{+}$ 
with the momentum of 1.3~GeV/$c$ (Eq.~(\ref{eq:wcnpe})).
As shown in the figure,
{\cerenkov} radiation yields more photons of shorter wavelength
because the number of 
photons is inversely proportional to the wavelength ($N$ in Eq.~(\ref{eq:wcnpe})).
Therefore, it is a key point
to minimize light absorption particularly for 
ultraviolet light in order to achieve larger NPE.
Light loss is mainly caused by the water, reflection material,
acrylic window, optical coupling grease, and PMT window glass. 
The quantum efficiency of PMT should be also considered 
as it is related to the detection of {\cerenkov} light.
Thus, a figure of merit (FoM) is defined as follows:
\begin{eqnarray} 
  \label{eq:fom_npe}
  {\rm FoM} &=& \int^{\infty}_{0}
  F(\lambda) d\lambda \nonumber \\
  &=& \int^{\infty}_{0}
  \Bigl( \epsilon_{{\rm water}}(\lambda) \times 
  \epsilon_{{\rm window}}(\lambda) \times 
  \epsilon_{{\rm grease}}(\lambda)  \nonumber \\
  && \times \epsilon_{{\rm pmt\_window}}(\lambda) \times 
  \epsilon^{\prime}_{{\rm pmt\_qe}}(\lambda) \Bigr) d\lambda,
\end{eqnarray}
where $\epsilon(\lambda)$ is the light-transmitting efficiency for 
each material and 
$\epsilon^{\prime}_{{\rm pmt\_qe}}(\lambda)$ is the PMT quantum efficiency. 
Here, the absorption by the reflection material is neglected as the 
wavelength dependence of the light absorption is nearly flat 
in the range we consider.
We attempted to determine the best configuration that maximizes the FoM 
with our choice of materials, taking into account the cost.

\begin{figure}[!htbp]
  \begin{center}
    \includegraphics[width=8.6cm]{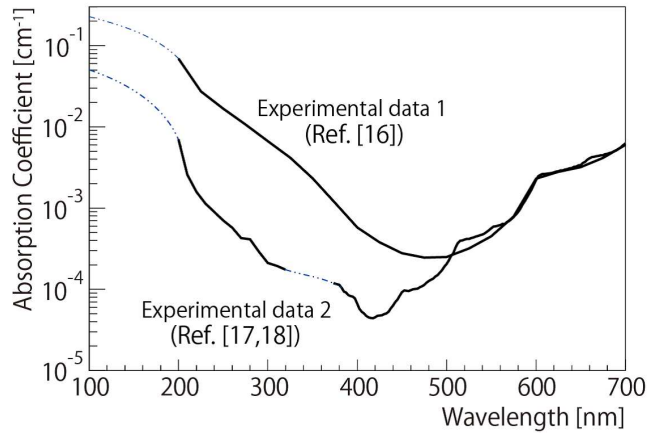}
    \caption{Absorption spectrum of water. Experimental data were 
      taken from~\cite{cite:water1,cite:water2,cite:water3}.
      Dashed lines represent extrapolations and an interpolation of the experimental data. 
      The internal transmittances are approximately $1\%$ and $99\%$ for 
      the absorption coefficients of $10^{-1}$ and $10^{-3}$~[/cm], respectively 
      if the path length of {\cerenkov} light is $\frac{35.0}{\cos{\theta_{C} (\beta=1)}}$~cm. 
    }
    \label{fig:h2o_absorption}
  \end{center}
\end{figure}

\begin{figure}[!htbp]
  \begin{center}
    \includegraphics[width=8.6cm]{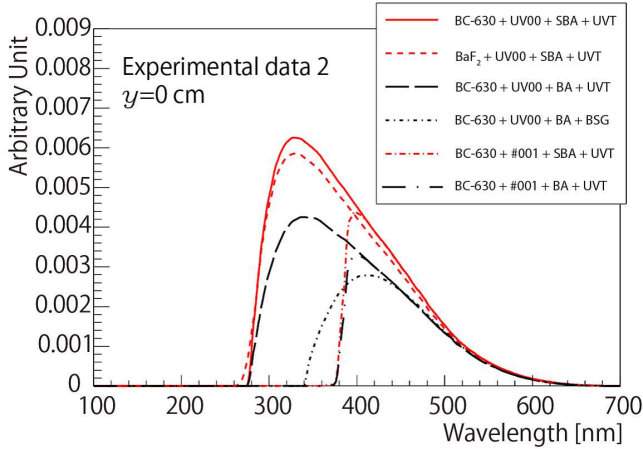}
    \caption{$F(\lambda)$ (Eq.~(\ref{eq:fom_npe})) with several
      configurations at $y=0$ using Experimental data~2.
      The ordinate axis was normalized so as to 
      make the FoM with a configuration of (BC-630+UV00+SBA+UVT) unity.} 
    \label{fig:npe_arbi}
  \end{center}
\end{figure}

\paragraph{Light Absorption by Water}
Examples of water-absorption spectra are shown in Fig.~\ref{fig:h2o_absorption}.
In the figure, three experimental data were plotted. 
One is taken from \cite{cite:water1} and is labeled as 
``Experimental data~1''.
The others are taken from \cite{cite:water3} and \cite{cite:water2} 
with covered ranges of 200--320~nm and 380--700~nm, respectively.
These two data are labeled as ``Experimental data~2''.
Dashed lines represent extrapolations and an interpolation of the 
experimental data. 
Although the extrapolation and interpolation of data 
would not be valid, we used them for the calculation of $F(\lambda)$ 
as shown later (Fig.~\ref{fig:npe_arbi}).
The absorption coefficients, especially for the UV region, 
are quite different between ``Experimental data 1'' and 
``Experimental data~2''.
The difference is caused by 
a difference of water purity, and purer water has 
a smaller absorption coefficient for the UV region~\cite{cite:water3}.
Thus, the water used in
the water {\cerenkov} detector should be as pure as possible,
and the water should be handled carefully to avoid any contamination.

\begin{figure}[!htbp]
  \begin{center}
    \includegraphics[width=8.6cm]{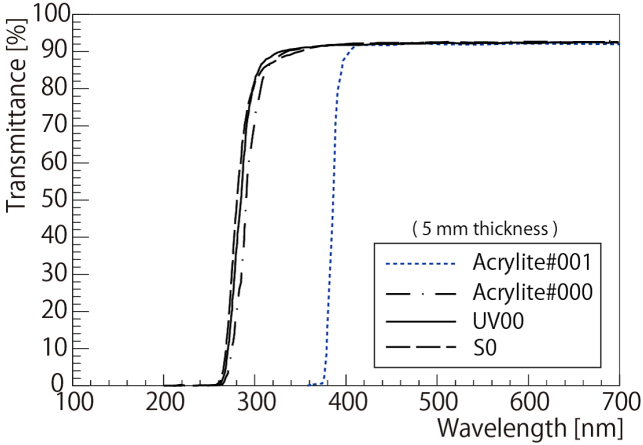}
    \caption{Transmittances of acrylics 
      (Acrylite$\#001$, Acrylite$\#000$, UV00, S-0) with thicknesses of 5~mm
      measured using a spectrophotometer (Shimazu Corporation MPS-2000).}
    \label{fig:acrylic_trans}
  \end{center}
\end{figure}

\paragraph{Transmittance of Acrylic Window}
We measured the transmittances of 
the following acrylic materials with thicknesses of 5~mm 
by using a spectrophotometer (Shimazu Corporation MPS-2000):
\begin{itemize}
\item Acrylite$\#$001
\item Acrylite$\#$000\footnote{Mitsubishi Rayon company, https://www.mrc.co.jp/}
\item UV00\footnote{Kuraray company, http://www.kuraray.co.jp/} 
\item S-0\footnote{Nitto Jushi Kogyo company, http://www.clarex.co.jp/}
\end{itemize}
The results are shown in Fig.~\ref{fig:acrylic_trans}.
The transmittance of Acrylite$\#001$, which was used for 
the previous prototype water {\cerenkov} detector (Sec.~\ref{sec:old_pwc}),
drops at approximately 380~nm.
On the other hand, the transmittances of other acrylics drop 
at approximately 280~nm.

\begin{figure}[!htbp]
  \begin{center}
    \includegraphics[width=8.6cm]{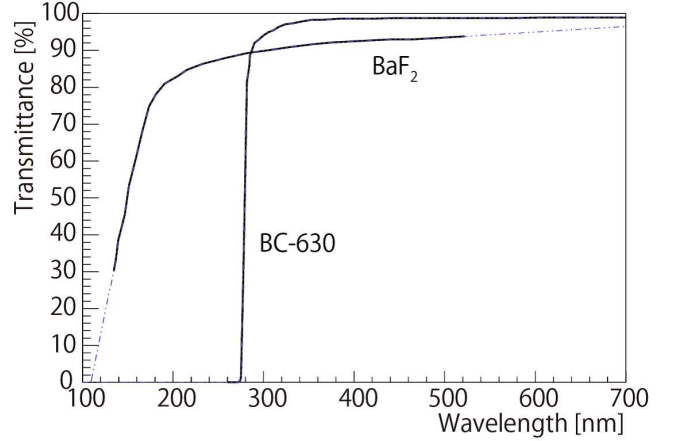}
    \caption{Transmittances of Saint-Gobain BC-630 (thickness of 0.114~mm)~\cite{cite:bc630} 
      and OKEN BaF$_{2}$ (thickness of 3~mm).
      Dashed lines for BaF$_{2}$ represent extrapolations of the data.}
    \label{fig:grease_trans}
  \end{center}
\end{figure}

\paragraph{Transmittance of Optical Coupling Grease}
The transmittance of an optical coupling grease, 
Saint-Gobain BC-630 ($n=1.465$), which was taken from
\cite{cite:bc630}, is shown in Fig.~\ref{fig:grease_trans}.
For comparison, 
the nominal transmittance of 
an optical crystal, OKEN\footnote{Ohyo Koken Kogyo company, http://www.oken.co.jp/web$\_$oken/indexen.htm} 
BaF$_{2}$ ($n=1.48$\footnote{This value is at the wavelength of 400~nm.}),
is also shown (dashed lines for BaF$_{2}$ represent extrapolations of the data).

\begin{figure}[!htbp]
  \begin{center}
    \includegraphics[width=8.6cm]{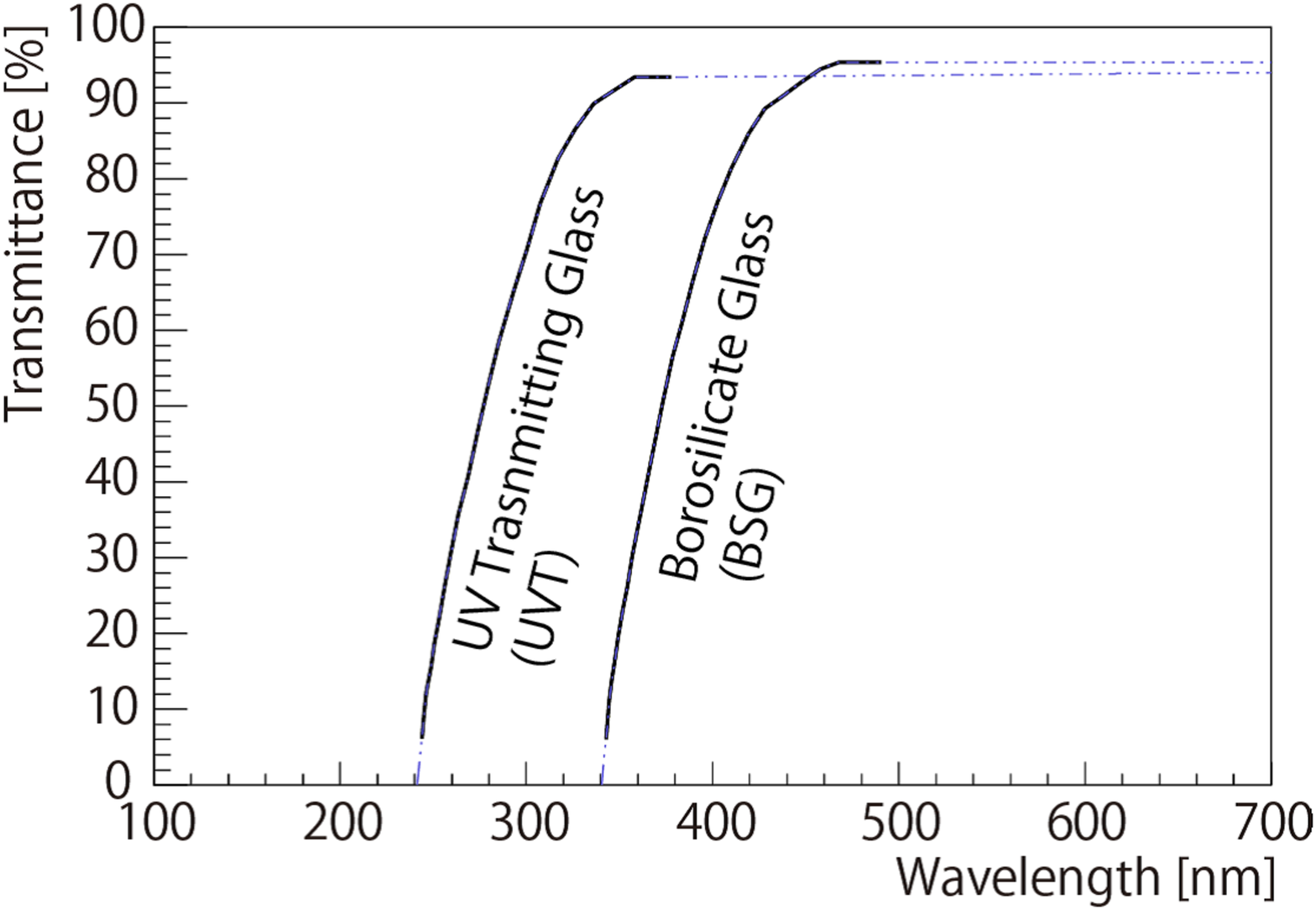}
    \caption{Nominal transmittances of PMT windows 
      (UV transmitting glass (UVT), borosilicate glass (BSG)
      )~\cite{cite:hamamatsu}.
      Dashed lines represent extrapolations of the data.}
    \label{fig:pmtwindow_trans}
  \end{center}
\end{figure}

\paragraph{Transmittance of PMT Window}
Before {\cerenkov} light reaches a PMT photocathode, 
the light has to pass through a PMT window. 
The nominal transmittances of PMT windows 
(UV transmitting glass (UVT), borosilicate glass (BSG)) 
are shown in Fig.~\ref{fig:pmtwindow_trans}~\cite{cite:hamamatsu}
(dashed lines represent extrapolations of the data).
The transmittance of the UVT is higher in the UV region compared to that of BSG.

\begin{figure}[!htbp]
  \begin{center}
    \includegraphics[width=8.6cm]{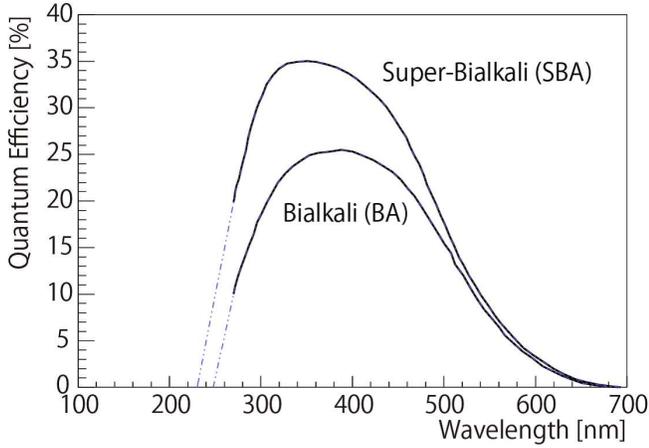}
    \caption{Quantum efficiencies of PMT photocathodes, 
      bialkali(BA) and super bialkali (SBA)~\cite{cite:hamamatsu}.
      Dashed lines represent extrapolations of the data.}
    \label{fig:pmt_qe}
  \end{center}
\end{figure}

\paragraph{PMT Quantum Efficiency}
Fig.~\ref{fig:pmt_qe} shows the nominal quantum efficiencies of 
PMT photocathodes (bialkali (BA) and super bialkali (SBA))~\cite{cite:hamamatsu}.
The quantum efficiency of SBA is higher than that of BA by 
approximately $10\%$ at the wavelength of 350~nm
(dashed lines represent extrapolations of the data).

\paragraph{Search for the Best Configuration}
The $F(\lambda)$ in Eq.~(\ref{eq:fom_npe}) for a different choice of the above materials was calculated, and 
some of results at $y=0$~cm using Experimental data~2 are shown in Fig.~\ref{fig:npe_arbi}.
The results were calculated for a 
$\pi^{+}$ with a momentum of 1.3~GeV/$c$, assuming 
that the path length of generated {\cerenkov} light 
in the container is $\frac{ 35.0 }{ \cos{\theta_{C}} }$~cm.
The ordinate axis was normalized so as to 
make the FoM with a configuration of (BC-630+UV00+SBA+UVT) unity.
FoM calculation results for some typical configurations 
are listed in Table~\ref{tab:exp_npe1}.
It is noted that the FoM with 
configuration 1 (BC-630+UV00+SBA+UVT) using 
Experimental Data~2 is normalized to unity in Fig.~\ref{fig:npe_arbi}.
\begin{table*}[!htbp]
  \begin{center}
    \caption{Normalized FoM for some typical configurations.
      The FoM with the configuration of 1 (BC-630+UV00+SBA+UVT) using 
      Experimental data~2 is normalized to unity.
      UV00, S-0 and $\#000$ are the acrylic windows
      shown in Fig.~\ref{fig:acrylic_trans}. 
    }
    \label{tab:exp_npe1}
    \begin{tabular}{|c|c|c|} 
      \hline \hline
      Configuration  & Data of & Normalized FoM \\ 
      & Water Absorption&  \\ \hline \hline
      1) & 1 &  0.916  \\ \cline{2-3}
      BC-630+UV00+SBA+UVT& 2 & 1  \\ \hline
      2) & 1 &  0.866\\ \cline{2-3}
      BaF$_{2}$+UV00+SBA+UVT& 2 &  0.948 \\ \hline
      3) & 1 & 0.868 \\ \cline{2-3}
      BaF$_{2}$+S-0+SBA+UVT& 2 &  0.952  \\ \hline
      4) & 1 &  0.848 \\ \cline{2-3}
      BaF$_{2}$+$\#$000+SBA+UVT& 2 & 0.923 \\ \hline
      5) & 1 &  0.675 \\ \cline{2-3}
      BC-630+UV00+BA+UVT& 2 &  0.73 \\ \hline
      6) & 1 &  0.396 \\ \cline{2-3}
      BC-630+UV00+BA+BSG& 2 &  0.404 \\ \hline
      7) & 1 &  0.438 \\ \cline{2-3}
      BC-630+$\#$000+SBA+UVT& 2 &  0.444 \\ \hline
      8) & 1 &  0.352 \\ \cline{2-3}
      BC-630+$\#$000+BA+UVT& 2 &  0.356 \\ \hline \hline
    \end{tabular}
  \end{center}
\end{table*}
The prototype water {\cerenkov} detector 
used for the positron beam test (Sec.~\ref{sec:old_pwc}) 
adopted configuration 7 (BC-630+$\#001$+SBA+UVT)
in Table~\ref{tab:exp_npe1}.
If one changes the acrylic window (Acrylite$\#001$) to 
one transmitting UV light such as UV00 (configuration 1),
the NPE is expected to be significantly improved compared to
that of the previous configuration. 
This significant improvement is caused by a 
recovery of the light spectrum in the UV region, as shown in Fig.~\ref{fig:npe_arbi}.
In order to confirm the above estimation and 
determine a {\cerenkov} light window material maximizing NPE, 
we performed a cosmic-ray test, which is described in the next section.

\subsection{Cosmic-ray Test for Optimizing the Window Material}
\label{sec:exp_setup_window}
\begin{figure}[!htbp]
  \begin{center}
    \includegraphics[width=8.6cm]{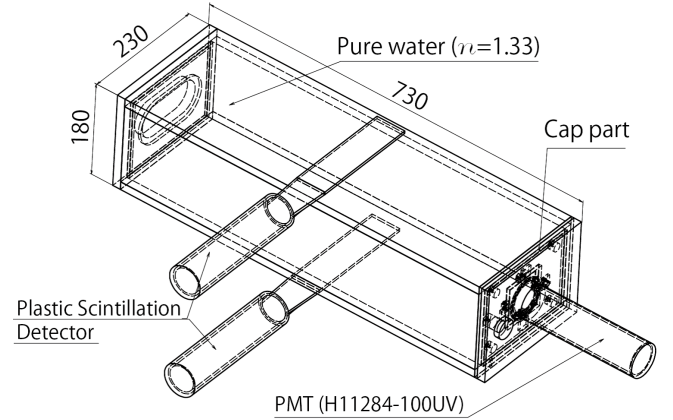}
    \caption{Schematic drawing of the experimental setup of 
      a cosmic-ray test to optimize the acrylic window material.
      The dimensions are in mm.}
    \label{fig:window_setup}
  \end{center}
\end{figure}

Figure~\ref{fig:window_setup} shows a schematic drawing of the experimental setup for 
optimizing the window material using cosmic rays. 
The container was the same as that used in the beam test (Sec.~\ref{sec:old_pwc}), 
but a cap part was newly designed and constructed in order to 
change the acrylic windows easily. 
A PMT of H11284-100UV was attached on the UVT acrylic window by using BaF$_{2}$.
The cap part was attached to the container using a PP band, 
and the container was filled with pure water. 
An LED was attached on the PMT near the photocathode
for use in the NPE calibration, as described in Sec.~\ref{sec:led_calib}.
The container was sandwiched between 
two plastic scintillation counters, which were 
used as data-taking triggers. 
The plastic scintillation detectors were set at $y=0$~cm during 
this test.
In the test, NPEs with UVT acrylic materials of 
UV00, S-0, and Acrylite$\#000$ (Fig.~\ref{fig:acrylic_trans}) were studied 
and compared with that resulting from Acrylite$\#001$.

\begin{figure}[!htbp]
  \begin{center}
    \includegraphics[width=8.6cm]{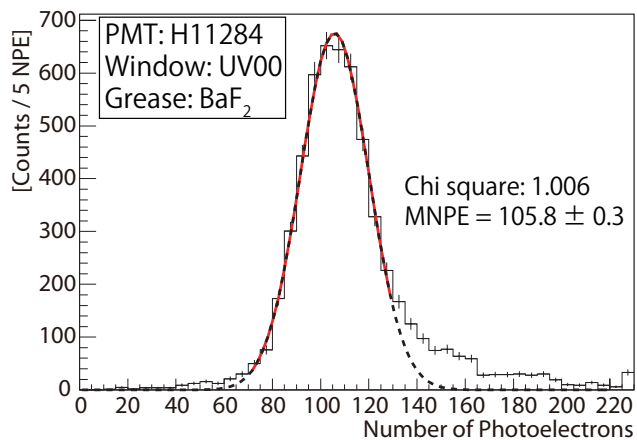}
    \caption{Typical NPE histogram obtained in a test to optimize 
      the acrylic window material. The NPE histogram was fitted 
      by a single-Gaussian function, and the mean value was used as the result.}
    \label{fig:npe_windowtest}
  \end{center}
\end{figure}

A conversion from ADC to NPE was performed using data 
with LED light, as described in Sec.~\ref{sec:led_calib}.
A typical NPE histogram is shown in Fig.~\ref{fig:npe_windowtest}.
The NPE spectrum was fitted with 
a single-Gaussian function to obtain the mean value (MNPE),
which was used as the optimization parameter in this cosmic-ray test. 

The MNPE results for UV00, S-0, and Acrylite$\#000$ are 
summarized in Table~\ref{tab:results_windowtest}.
Errors in the results are statistical and systematic 
($=\sigma_{calb.}$, Sec.~\ref{sec:sys-error}).
Each acrylic window was tested twice to check
the reproducibility, and the MNPEs of two measurements 
for each window usage were consistent within the errors.
All the tested windows were able to achieve 
better MNPEs than that of Acrylite$\#001$ by $\geq50\%$.
The NPEs with UV00 and S-0 were larger than that of 
Acrylite$\#000$ by $\geq5\%$, though a difference of 
NPEs between UV00 and S-0 was not observed within the errors. 
Consequently, we decided to use UV00 as the {\cerenkov} 
window material. 
\begin{table}[!htbp]
  \begin{center}
    \caption{Mean NPE (MNPE) results obtained using  
      UV00, S-0, and Acrylite$\#000$ as the acrylic window material. Each window was tested 
      twice to check the reproducibility of the MNPE measurement. 
      Errors in the results are statistical and 
      systematic ($=\sigma_{calb.}$, Sec.~\ref{sec:sys-error}).}
    \label{tab:results_windowtest}
    \begin{tabular}{|c|c|c|}
      \hline \hline
      Window material & MNPE & Trial \\ \hline
      UV00 & $105.8 \pm 0.3 \pm 1.1$ & 1\\
           & $103.9 \pm 0.6 \pm 1.0$ & 2\\ \hline 
      S-0  & $106.6 \pm 0.3 \pm 1.1$ & 1\\
           & $105.4 \pm 0.6 \pm 1.1$ & 2\\ \hline 
      Acrylite$\#$000 & $96.4 \pm 0.4 \pm 1.0$ & 1\\
                      & $99.6 \pm 0.5 \pm 1.0$ & 2\\
      \hline \hline
    \end{tabular}
  \end{center}
\end{table}

It is noted that BaF$_{2}$ was used as 
an optical coupling grease in this test. 
We also measured MNPE with BC-630 instead of BaF$_{2}$, 
but no difference was observed.
However, BaF$_{2}$ has the lower viscosity than that of BC-630.
In terms of stability, 
we decided to use BC-630. 

\section{New Prototype Water {\cerenkov} Detector}
The window material was optimized
as described in the previous section. 
A new prototype was constructed with the new window material, 
which has high transmittance in the UV region. 
In this section, the design of the new prototype and 
the performance test with cosmic rays are described.

\subsection{Design}
\begin{figure}[!htbp]
  \begin{center}
    \includegraphics[width=8.6cm]{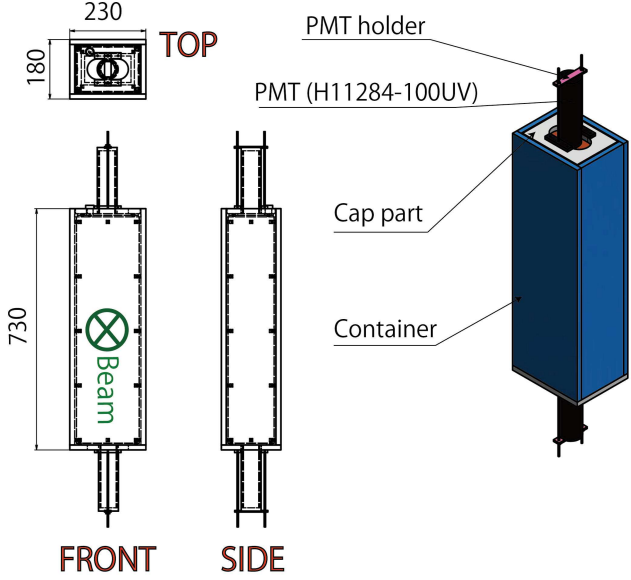}
    \caption{Drawing of the new prototype water {\cerenkov} detector 
      constructed with the optimized window material (UV00).
      The dimensions are in mm.}
    \label{fig:newwc1}
  \end{center}
\end{figure}

\begin{figure}[!htbp]
  \begin{center}
    \includegraphics[width=8.6cm]{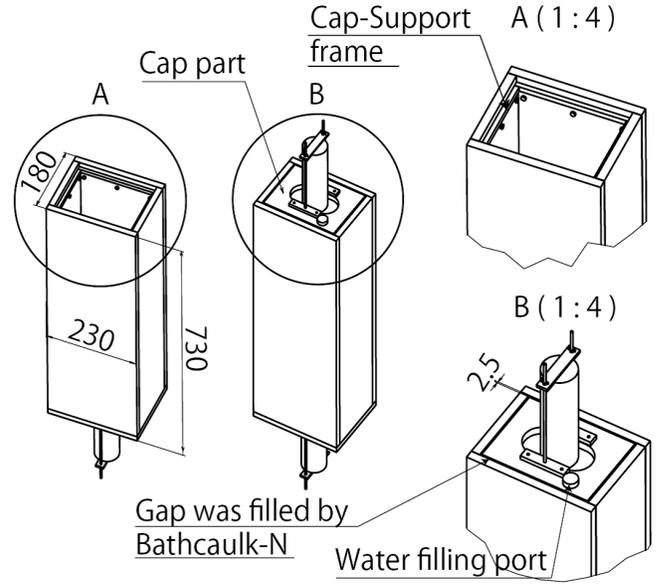}
    \caption{Design of cap part of new prototype water {\cerenkov} detector.
      The dimensions are in mm.}
    \label{fig:wc_cappart}
  \end{center}
\end{figure}

The design of the 
new prototype water {\cerenkov} detector
is similar to that of the previous prototype, as described in 
Sec.~\ref{sec:prototype_design0}, except for 
window material, cap part and the way to attach the reflection material 
inside the container.
Figure~\ref{fig:newwc1} shows a drawing of the 
new prototype water {\cerenkov} detector.
UV00 was chosen (Sec.~\ref{sec:exp_setup_window}) 
and used as the window material, which was bonded
to an inner surface of the container box by polymerization.
The cap part was designed as shown in Fig.~\ref{fig:wc_cappart}.
The size of the cap part is smaller than 
the opening mouth of the container box. 
There is a frame on the inner surface of the container to support the cap.
Then, a caulking material (CEMEDINE, 
Bathcaulk N\footnote{CEMEDINE Co., http://www.cemedine.co.jp/e/index.html}) 
was filled into a gap between the cap part and container box
to seal the detector.
Small plastic pieces which have a screw hole were 
bonded to the inner surface of the container box by polymerization,  
and Tyvek sheets (reflection material) were 
attached to the plastic pieces by plastic screws.
The effective volume was $190^{W} \times 690^{H} \times 140^{T}$~mm$^{3}$.
\subsection{Cosmic-ray Test}
In the cosmic-ray test, the dependence of mean NPE (MNPE) on the 
incident particle position, applied HV, and 
time after construction were measured. 
Furthermore, systematic errors in NPE due to calibration, 
difference in individual PMT performance, and the effect of the 
optical coupling grease were investigated. 

\subsubsection{Experimental setup}
\begin{figure}[!htbp]
  \begin{center}
    \includegraphics[width=8.6cm]{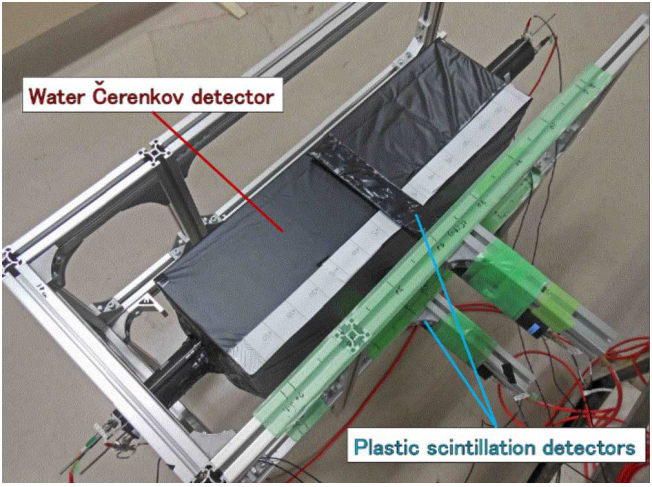}
    \caption{Photograph of the experimental setup of 
      the cosmic-ray test for the prototype water {\cerenkov} detector.}
    \label{fig:picture_cosmicray_test}
  \end{center}
\end{figure}

The experimental setup was similar to that 
of the cosmic-ray test described in Sec.~\ref{sec:exp_setup_window}.
The prototype water {\cerenkov} detector 
was sandwiched by two plastic scintillation detectors, 
which were used as trigger counters for data collection. 
The two trigger counters could be moved 
to observe the dependence of mean NPE on the incident particle position.
Fig.~\ref{fig:picture_cosmicray_test} shows a 
photograph of the experimental setup.
An LED, which was used for the NPE calibration, was attached
beside a PMT photocathode\footnote{Outside of the container box.}. 
\subsubsection{LED Calibration and Systematic Error}
\label{sec:sys-error}
\begin{figure}[!htbp]
  \begin{center}
    \includegraphics[width=8.6cm]{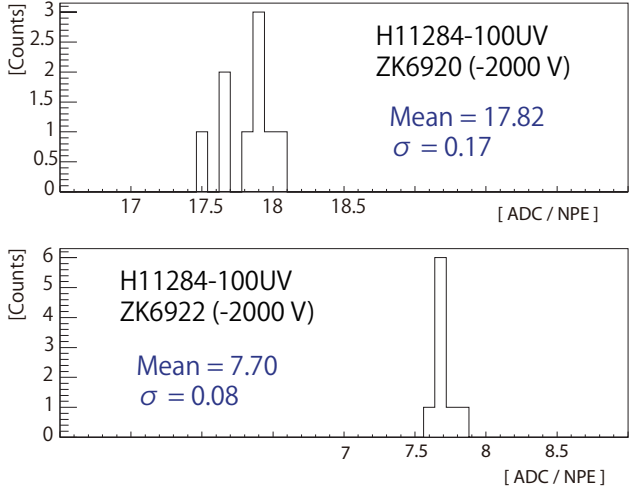}
    \caption{Nine of the obtained LED calibration data 
      for H11284-100UV PMTs with serial numbers of ZK6922 ($-2000$~V) and ZK6920 ($-2000$~V)
      in a period of one week.}
    \label{fig:led_fluctuation}
  \end{center}
\end{figure}

The ADC channels for a single photoelectron 
is obtained by fitting an ADC histogram with LED 
as shown in Fig.~\ref{fig:wc1LED}.
Possible fluctuations of the calibration process due to temperature {\it etc.} 
were examined by repeating the calibration measurements for two different 
PMTs in a period of one week.
The obtained ADC channels per photoelectron for 
H11284-100UV PMTs with serial numbers of ZK6920 ($-2000$~V) and ZK6922 ($-2000$~V)
are plotted in Fig.~\ref{fig:led_fluctuation}.
It is found that the standard deviations were $<1\%$ for both PMTs. 
In the present report, 
the systematic error of the resulting NPE is taken as $1\%$ ($\equiv \sigma_{calb.}$).

\subsubsection{NPE Detection Performance}
\label{sec:npe_newprototype}
The mean NPEs (MNPE) detected by TOP (ZK6922, $-2000$~V) and BOTTOM (ZK6920, $-2000$~V) 
PMTs at $y=0$~cm\footnote{Coordinates are defined as Fig.~\ref{fig:wc_onebox}.} were 
\begin{eqnarray}
  {\rm MNPE}^{{\rm Cosmic}}_{({\rm TOP,ZK6922})} = 109.6 \pm 0.5^{stat.} \pm 1.1^{sys.}, \\
  {\rm MNPE}^{{\rm Cosmic}}_{({\rm BOT,ZK6920})} = 97.9 \pm 0.5^{stat.} \pm 1.1^{sys.}. 
\end{eqnarray}
The mean summed NPE (MsNPE) obtained by fitting to a summed NPE histogram was
\begin{eqnarray}
  {\rm MsNPE}^{{\rm Cosmic}} = 207.7 \pm 0.9^{stat.} \pm 2.9^{sys.} .  
\end{eqnarray}
The difference between TOP and BOTTOM MNPEs mainly originates
from how the Tyvek sheet was attached inside the container, as 
discussed quantitatively in Sec.~\ref{sec:pmt_ind_diff}.

\subsubsection{$y$-Position Dependence}
\label{sec:posdep_npe}
\begin{figure}[!htbp]
  \begin{center}
    \includegraphics[width=8.6cm]{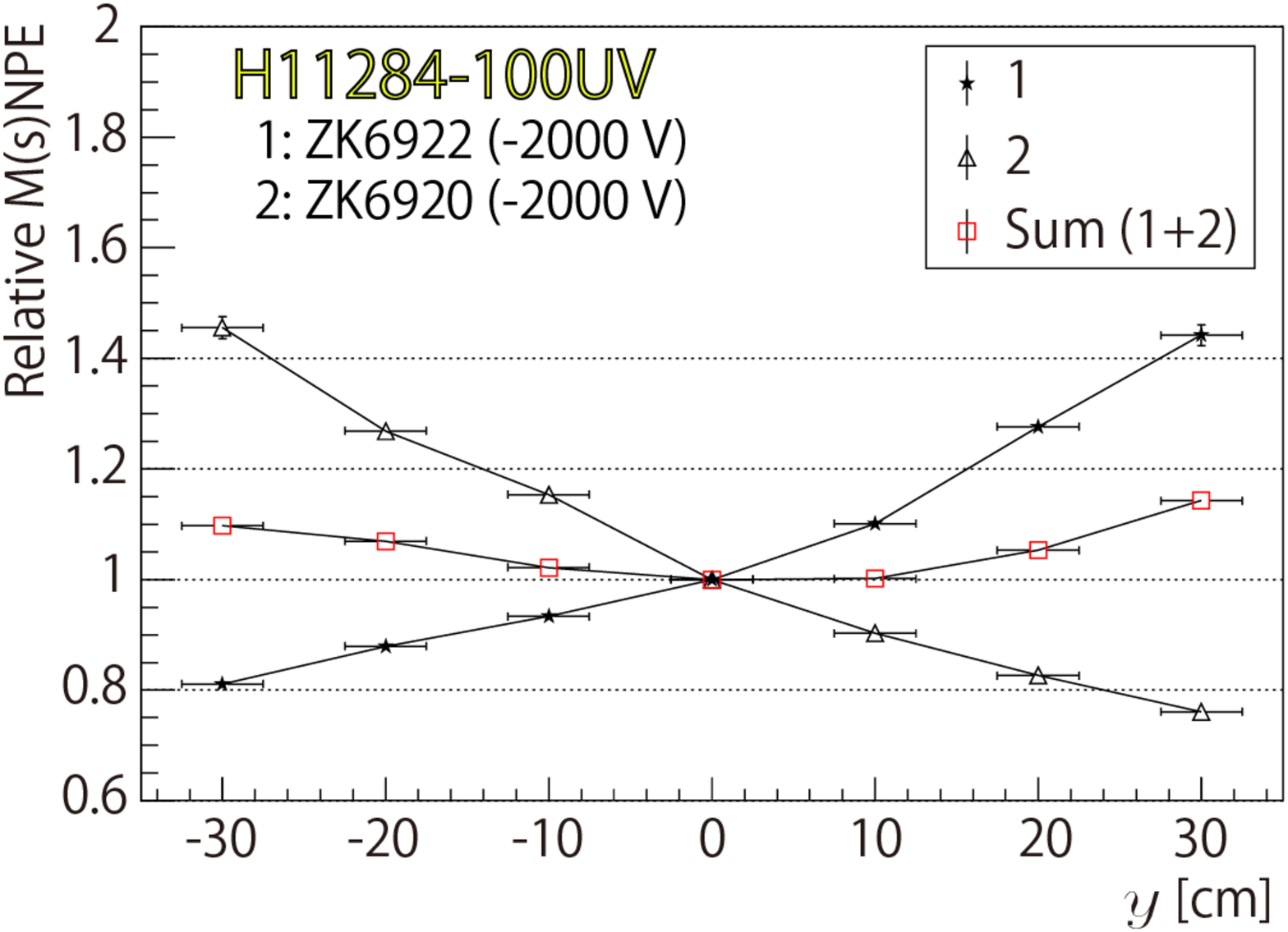}
    \caption{$y$-position dependence of the MNPE for TOP/BOTTOM (labeled as 1/2) PMT
      as well as the MsNPE.}
    \label{fig:ydep_real}
  \end{center}
\end{figure}

The dependence of MNPE on incident particle positions 
is rather stronger in the $y$-direction than in the $x$-direction, 
as shown in the positron beam test (Fig.~\ref{fig:xydep_2d}).
Thus, the $y$-position dependence was investigated by changing 
the position of trigger scintillation counters in the $y$-direction. 
Figure~\ref{fig:ydep_real} shows resulting M(s)NPEs as a function 
of the $y$-position. The M(s)NPEs were normalized to unity at $y=0$~cm.
The $y$-position dependence is not flat for the summed NPE. 
The dependence might be caused by 
asymmetric tendencies of photon yield (Fig.~\ref{fig:np_wavelength}) and 
absorption coefficients of water (Fig.~\ref{fig:h2o_absorption})
for {\cerenkov} light with different wavelengths.
For {\cerenkov} light with a shorter wavelength, 
the light yield is greater, 
but the absorption coefficient 
of water is larger as well.
The MsNPE variation depending on the incident $y$-position 
was obtained as $<+20\%$.

\subsubsection{HV Dependence}
\begin{figure}[!htbp]
  \begin{center}
    \includegraphics[width=8.6cm]{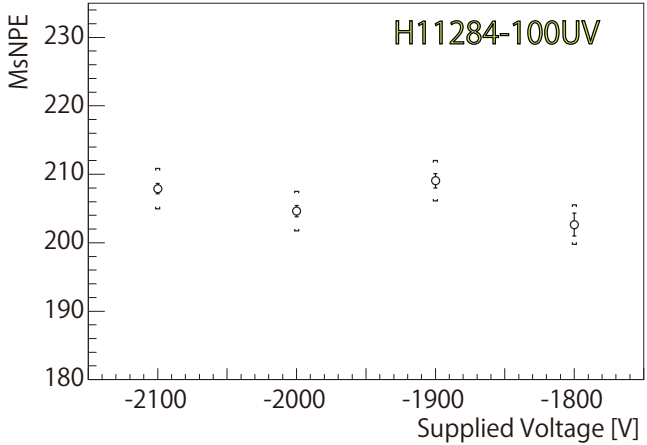}
    \caption{HV dependence of MsNPE.
      There is no dependence on HV within the errors.}
    \label{fig:hvdep}
  \end{center}
\end{figure}

The supplied high voltages (HVs) for PMTs were varied 
from $-1800$~V to $-2100$~V, and the HV dependence of NPE was investigated. 
Figure~\ref{fig:hvdep} shows the HV dependence of MsNPE
with statistical and systematic errors.
The systematic errors are represented by brackets 
($=\sigma_{calb.}$, Sec.~\ref{sec:sys-error}). 
No HV dependence was found within the errors. 
\subsubsection{Performance Difference of Individual PMTs}
\label{sec:pmt_ind_diff}
The NPE depends on the performance factors of individual PMTs, 
such as quantum efficiency, and 
this dependence was studied by measuring NPEs with PMTs of different serial numbers.
Table~\ref{tab:pmtdiff} lists 
serial numbers, applied HVs, ADC channels per photoelectron (ADC/NPE), and  
obtained MNPEs at $y=0$~cm.
\begin{table*}[!htbp]
  \begin{center}
    \caption{MNPE results using H11284-100UV PMTs 
      with different serial numbers at $y=0$~cm.
      Normalized MNPEs shown in the 
      last column were obtained using the geometrical 
      correction factor, $G_{T2B}$, which normalizes a TOP MNPE to a BOTTOM one.
      In this table, BOTTOM MNPEs were divided by $G_{T2B}$ 
      for comparison with each other.}
    \label{tab:pmtdiff}
    \begin{tabular}{|c|c|c|c|c|c|} 
      \hline \hline
      Serial & HV & ADC/NPE & PMT& MNPE & Normalized  \\ 
      number & [V] & & position & & MNPE\\ \hline
      ZK6919 & $-2250$ & 11.18& BOTTOM & $104.8 \pm 0.4 $& $121.9 \pm 0.4 $ \\
      ZK6920 & $-2000$ & 19.08 & BOTTOM & $97.9 \pm 0.5$ & $114.4 \pm 0.6$\\
      ZK6922 & $-2000$ & 7.21  & TOP    & $109.6 \pm 0.5$& $109.6 \pm 0.5$\\ 
      ZK6925 & $-2250$ & 7.04 & TOP    & $113.0 \pm 0.4 $& $113.0 \pm 0.4 $ \\
      ZK7295 & $-2000$ & 6.31 & BOTTOM & $101.4 \pm 1.0$ & $117.9 \pm 1.2$ \\
      ZK7296 & $-2000$ & 4.19 & BOTTOM & $102.1 \pm 0.5$ & $118.7 \pm 0.5$\\
      ZK7298 & $-2000$ & 9.97 & BOTTOM & $97.7 \pm 0.7$  & $113.6 \pm 0.8$\\
      ZK7299 & $-2000$ & 6.12 & TOP    & $112.2 \pm 0.6$ & $112.2 \pm 0.6$\\
      ZK7300 & $-2250$ & 7.63 & TOP    & $110.7 \pm 0.6$ & $110.7 \pm 0.6$\\
      ZK7301 & $-2250$ & 6.31 & TOP    & $126.7 \pm 0.4$ & $126.7 \pm 0.4$\\
      ZK7302 & $-2250$ & 7.87 & BOTTOM & $102.5 \pm 0.4$ & $119.2 \pm 0.4$\\
      ZK7303 & $-2250$ & 4.20 & BOTTOM & $113.5 \pm 0.4$ & $131.9 \pm 0.5$\\
      ZK7304 & $-2250$ & 5.21 & TOP    & $103.4 \pm 0.6$ & $103.4 \pm 0.6$\\
      ZK7305 & $-2250$ & 4.44 & BOTTOM & $109.4 \pm 0.7$ & $127.2 \pm 0.8$\\
      ZK7306 & $-2250$ & 3.80 & TOP    & $117.1 \pm 0.5$ & $117.1 \pm 0.5$ \\
      ZK7307 & $-2250$ & 13.50& TOP    & $124.7 \pm 1.0$ & $124.7 \pm 1.0$\\
      ZK7308 & $-2250$ & 11.82 & TOP   & $115.8 \pm 0.3$ & $115.8 \pm 0.3$\\
      ZK7309 & $-2250$ & 7.71 & BOTTOM & $100.3 \pm 0.7$ & $116.7 \pm 0.8$\\
      ZK7310 & $-2250$ & 4.58 & BOTTOM & $98.7 \pm 0.6$  & $114.7 \pm 0.7$\\
      ZK7311 & $-2000$ & 10.88& BOTTOM & $94.4 \pm 0.5$  & $109.8 \pm 0.5$\\
      ZK7312 & $-2250$ & 6.46 & TOP    & $114.0 \pm 0.5$ & $114.0 \pm 0.5$\\
      ZK7313 & $-2000$ & 7.81 & TOP    & $111.1 \pm 0.4$ & $111.1 \pm 0.4$\\
      ZK7315 & $-2250$ & 8.84 & TOP    & $112.2 \pm 0.4$ & $112.2 \pm 0.4$\\
      ZK7316 & $-2250$ & 3.31 & TOP    & $116.3 \pm 0.5$ & $116.3 \pm 0.5 $ \\
      \hline \hline
    \end{tabular}
  \end{center}
\end{table*}

\begin{figure}
  \begin{center}
    \includegraphics[width=8.6cm]{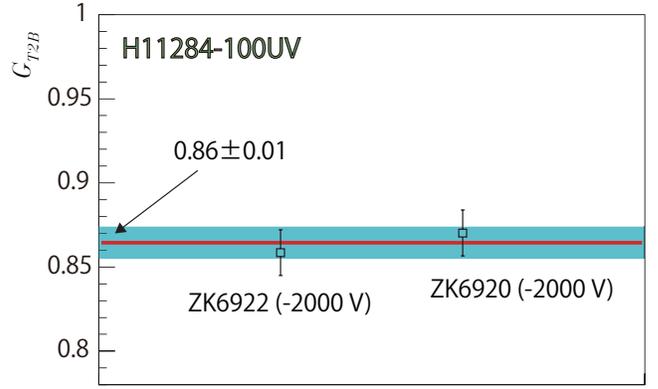}
    \caption{BOTTOM MNPE ratios relative to the TOP ones for the PMTs 
      ZK6922 and ZK6920. The fitting result is represented by
      a line, and the geometrical correction factor was obtained as 
      $G_{T2B}=0.86$ ($\pm 0.01$).}
    \label{fig:geof}
  \end{center}
\end{figure}

There is a systematic difference between MNPEs 
detected by TOP and BOTTOM PMTs.
The container geometry was not symmetric, because 
the cap part was only attached to the TOP side.
The geometric asymmetry of the container 
made the reflection-sheet attachments asymmetric,
and we observed a systematic NPE difference between 
TOP and BOTTOM MNPEs.
In order to obtain a geometrical correction factor ($\equiv G_{T2B}$)
that normalizes a TOP MNPE to that of BOTTOM, 
data swapping of TOP and BOTTOM PMTs was performed. 
Table~\ref{tab:pmt_swap} lists MNPE results of the swapping test.
\begin{table}[!htbp]
  \begin{center}
    \caption{MNPE results with H11284-100UV PMTs of ZK6922
      and ZK6920 when they were attached 
      on the TOP and BOTTOM sides. 
      The errors are statistical.}
    \label{tab:pmt_swap}
    \begin{tabular}{|c|cc|}
      \hline \hline
      &  ZK6922 ($-2000$~V) & ZK6920 ($-2000$~V)\\ \hline
      TOP &  $109.6 \pm 0.5$ & $112.5 \pm 0.5$\\
      BOTTOM & $94.1 \pm 0.5$ & $97.9 \pm 0.3$\\ \hline \hline
    \end{tabular}
  \end{center}
\end{table}
BOTTOM MNPE ratios relative to TOP ones are shown in Fig.~\ref{fig:geof}. 
The geometrical correction factor was obtained as 
$G_{T2B}=0.86$ $(\pm 0.01)$ through a fitting to the results, as represented 
by a line in Fig.~\ref{fig:geof}. 

\begin{figure}[!htbp]
  \begin{center}
    \includegraphics[width=8.6cm]{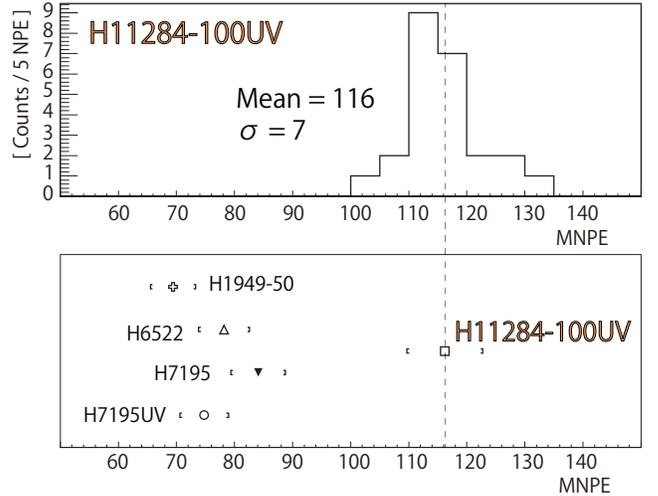}
    \caption{(Top) Histogram filled with
      center values of normalized MNPEs by $G_{T2B}$
      with H11284-100UV PMTs (see also Table~\ref{tab:pmtdiff}). 
      (Bottom) Comparison of MNPEs among different types of PMTs.}
    \label{fig:nnpe_filled}
  \end{center}
\end{figure}

MNPEs normalized using $G_{T2B}$ 
for comparison with each other are listed in 
the last column of Table~\ref{tab:pmtdiff}. 
The BOTTOM MNPEs were normalized to those of 
TOP in the table (BOTTOM MNPEs were divided by $G_{T2B}$).
The center values of normalized MNPEs were filled in a histogram 
and shown in the top panel in Fig.~\ref{fig:nnpe_filled}.
The mean value and standard deviation
were obtained as 116 and 7 ($=6\%$), respectively.
The $6\%$ standard deviation also includes errors 
due to the calibration ($\sigma_{calib.}=1\%$), 
reproducibility of PMT attachment, and statistics ($\leq 1\%$).
The errors of reproducibility are
expected to be smaller 
or comparable to that of the calibration ($\sigma_{calib.}$),
as indicated in Table~\ref{tab:results_windowtest} 
and Table~\ref{tab:grease_eff}. 
Thus, the performance difference of individual PMTs was 
estimated to be $\sqrt{6^{2}-1^{2}-1^{2}-1^{2}} \simeq 6\%$
($\equiv \sigma_{{\rm PMT}}$).

\subsubsection{PMT-Type Difference}
To confirm the PMT choice of H11284-100UV that 
has a super bialkali photocathode and UV transmitting 
window (SBA+UVT), other types of Hamamatsu PMTs 
(H6522, H1949-50, H7195, H7195UV)
were tested with cosmic rays. 
The MNPE results are listed in Table~\ref{tab:pmt-comparison} 
and the bottom panel of Fig.~\ref{fig:nnpe_filled}. 
Errors in the results are statistical and 
$\sigma_{{\rm PMT}}$\footnote{The systematic error ($\sigma_{calib.}$) is
negligibly small compared to $\sigma_{{\rm PMT}}$}.
It was found that the PMT H11284-100UV 
is able to achieve an MNPE greater than those achieved with the other types of 
PMTs by more than 30$\%$.
\begin{table*}[!htbp]
  \begin{center}
    \caption{MNPE measurement results
      of H11284-100UV (Sec.~\ref{sec:pmt_ind_diff}), 
      H6522 (BA+UVT), H1949-50 (BA+BSG), H7195 (BA+BSG), and H7195UV (BA+UVT). 
      BA and SBA stands for a type of photocathode as shown in Fig.~\ref{fig:pmt_qe}, and 
      UVT and BSG stands for a type of PMT window as shown in Fig.~\ref{fig:pmtwindow_trans}. 
      BOTTOM MNPEs were divided by $G_{T2B}$ for comparison with TOP ones.
      Errors in the results are statistical and $\sigma_{{\rm PMT}}$.}
    \label{tab:pmt-comparison}
    \begin{tabular}{|c|c|c|c|}
      \hline \hline
      \multicolumn{3}{|c|}{PMT} & Normalized  \\  \cline{1-3}
      Type & Photocathode + Window & Serial Number & MNPE \\ \hline 
      H11284-100UV & SBA + UVT & - & $116 \pm 7$ (Sec.~\ref{sec:pmt_ind_diff}) \\ 
      H6522 & BA + UVT & LA1537 & $78.2 \pm 0.3 \pm 4$\\
      H1949-50 & BA + BSG& WA6589 & $69.4 \pm 0.3 \pm 4$ \\
      H7195 & BA + BSG & RD7198 & $84.1 \pm 0.3 \pm 5$\\
      H7195UV & BA + UVT & LA1528& $74.8 \pm 0.3 \pm 4$\\ \hline \hline 
    \end{tabular}
  \end{center}
\end{table*}

\subsubsection{Effect of Optical Coupling Grease}
\label{sec:grease_eff}
To observe the NPE difference between the cases with and without 
the optical coupling grease, 
data without the use of the grease were taken and compared.
The test proceeded as follows: 
\begin{itemize}
\item The optical coupling grease (Saint-Gobain BC-630) was used between 
  the acrylic windows and PMTs for both the TOP and BOTTOM PMTs 
  in the first trial (Trial 1).
\item Then, the grease for the BOTTOM PMT was removed, and 
  the PMT and acrylic window were only physically attached (Trial 2). 
\item Finally, the grease was used for the BOTTOM PMT again 
  to check reproducibility (Trial 3).
\end{itemize}

Table~\ref{tab:grease_eff} shows the MNPE results.
The errors in the results are statistical and systematic ($\sigma_{calib.}$). 
\begin{table}[!htbp]
  \begin{center}
    \caption{MNPE results of the test to study the effect of the optical coupling grease
      (Saint-Gobain BC-630).}
    \label{tab:grease_eff}
    \begin{tabular}{|c|c|c|c|}
      \hline \hline
       Trial & ZK6922 ($-2000$~V) & ZK6920 ($-2000$~V) \\ 
             & [TOP] & [BOTTOM]  \\ \hline \hline
       1 & $109.6\pm0.5\pm1.1$ & $97.9\pm0.3\pm1.0$\\
         &                     & \\ \hline
       2 & $114.3\pm0.7\pm1.1$ & $63.6\pm0.3\pm0.6$\\ 
         &                     & (No Grease)  \\ \hline 
       3 & $111.6\pm0.5\pm1.1$ & $98.2\pm0.4\pm1.0$\\
         &                     &                   \\
       \hline \hline
    \end{tabular}
  \end{center}
\end{table}
It was found that the MNPE is reduced by approximately $35\%$ when
the optical coupling grease is not used. 
This reduction might be caused by the 
air gap between the PMT and acrylic window. 
If there is an air gap between a PMT window and 
acrylic window, total reflection 
occurs for a photon with a certain incident angle 
at the boundary.

\subsubsection{Long-Term Stability}
\begin{figure}[!htbp]
  \begin{center}
    \includegraphics[width=8.6cm]{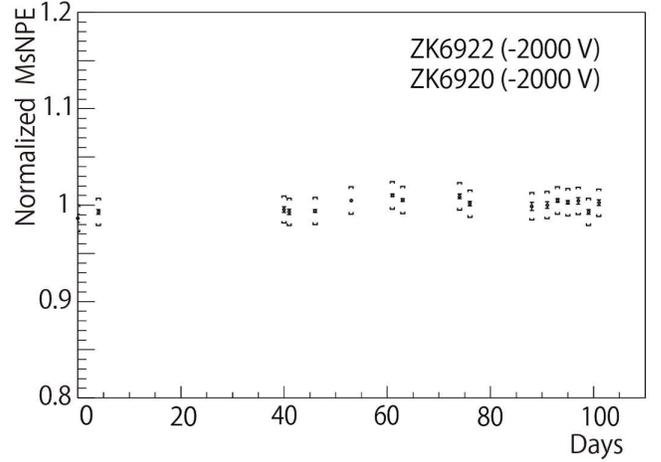}
    \caption{Time dependence of MsNPE in the cosmic-ray test.
      The prototype water {\cerenkov} detector was able to
      work stably for 100 days.}
    \label{fig:long_term_npe}
  \end{center}
\end{figure}

The water {\cerenkov} detector 
should work stably during the term of an experiment,
which is typically around a month.
In order to check long-term stability, 
NPE measurement was performed with cosmic rays for 100 days.
Figure~\ref{fig:long_term_npe} shows 
MsNPE results of the measurement.
The MsNPEs were normalized so as to 
make the mean value of obtained MsNPEs unity.
It was confirmed that the prototype water {\cerenkov} detector
was able to work stably for 100 days.

\section{Expected Proton Rejection Power by a Monte Carlo Simulation}
\label{sec:epr_geant}
\begin{figure}[!htbp]
  \begin{center}
    \includegraphics[width=8.6cm]{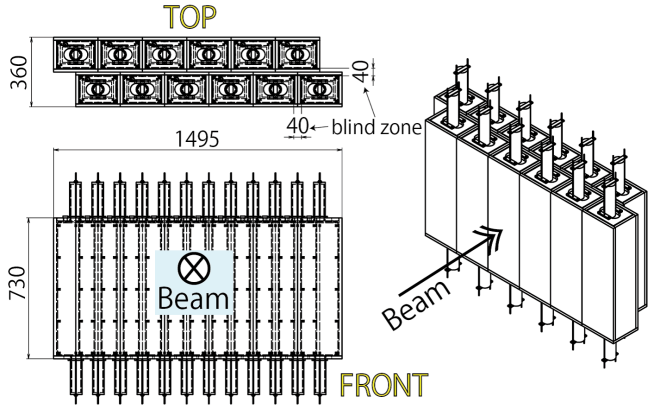}
    \caption{Schematic drawing of twelve segments of the water {\cerenkov} detectors.
      One layer consists of six segments. Two layers are 
      set to cover their inefficient regions from each other.
      The dimensions are in mm.
    }
    \label{fig:wc_12segments}
  \end{center}
\end{figure}

The performance of one segment of the water {\cerenkov} detector 
was discussed in the above sections.
In the experiment J-PARC E05, however, 
twelve segments of water {\cerenkov} detectors 
will be installed in two layers, as 
shown in Fig.~\ref{fig:wc_12segments}.
A Monte Carlo simulation (Geant4~\cite{cite:geant4}) was performed in order to 
study the performance of the water {\cerenkov} detector as 
a total system of twelve segments in S-2S. 
The expected performances of proton rejection efficiency and 
$K^{+}$ survival ratio obtained using the Monte Carlo simulation are 
described in this section.

\subsection{Setup and Conditions in the Monte Carlo Simulation}
\begin{figure}[!htbp]
  \begin{center}
    \includegraphics[width=8.6cm]{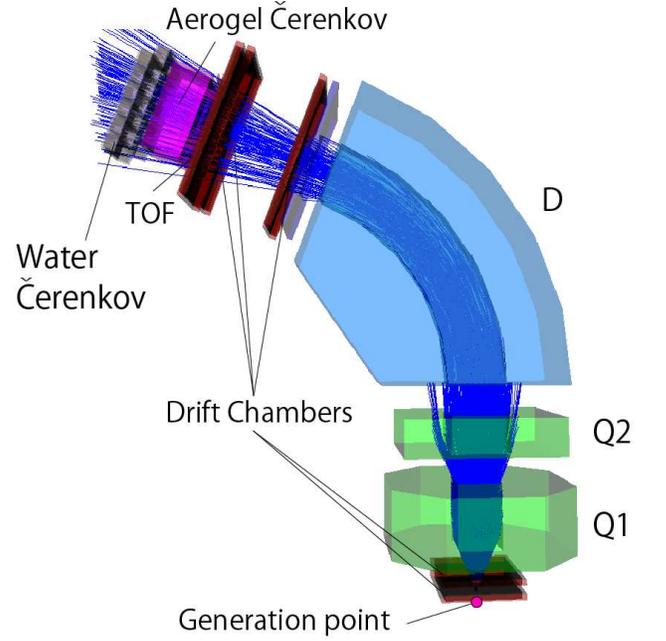}
    \caption{Displayed tracks in the S-2S Monte Carlo simulation.}
    \label{fig:s2s_sim}
  \end{center}
\end{figure}

In the simulation, S-2S is modeled as shown in Fig.~\ref{fig:s2s_setup}.
The magnetic field of S-2S was calculated 
with a software using the three-dimensional finite-element method, 
Opera3D (TOSCA)~\cite{cite:tosca}. 
Materials were defined for 
the S-2S magnets (QQD), TOF detector, and water {\cerenkov} detector
in the simulation\footnote{Others were set as vacuum.}.
The water {\cerenkov}
detector was placed 2.41~m downstream of  
the dipole magnet exit in the beam direction.
Particles were generated at a point 600~mm upstream
of the Q1 magnet entrance, where a target 
is planned to be installed.
The MsNPE in the water {\cerenkov} detector 
when a $\beta=1$ particle is perpendicularly incident at the center is 
assumed to be 200,
although the MsNPE was obtained to be $207.7 \pm 0.9 \pm 2.9$ 
in the cosmic-ray test, as described in Sec.~\ref{sec:npe_newprototype}.
The MsNPE value was randomly varied by $\pm 6\%$ event by event 
to take into account the performance difference of individual PMTs 
($\sigma_{{\rm PMT}}=6\%$) as discussed in Sec.~\ref{sec:pmt_ind_diff}.
The MsNPE position dependence, which was represented 
in Sec.~\ref{sec:posdep_npe}, was taken into account 
by scaling MsNPE according to the incident position of a particle. 
The path length of an incident particle in 
the medium of the water {\cerenkov} detector depends
on the incident position and angle. 
Thus, path-length scaling was applied
for the MsNPE calculation event by event (Eq.~(\ref{eq:wcnpe})). 
Fig.~\ref{fig:s2s_sim} shows 
particle tracks in the simulation.
Particles are made to converge sequentially in the longitudinal and transverse 
directions by using two quadrupole magnets (Q1 and Q2), and they are made to bend 
in a dispersive plane by using a dipole magnet (D) 
transporting the particles to the particle detectors. 
\begin{table}[!htbp]
  \begin{center}
    \caption{Conditions in the S-2S Monte Carlo simulation.}
    \label{tab:sim_condition}
    \begin{tabular}{|c|c|c|}
      \hline \hline 
      Particle Generation & Distribution& Spherical uniform \\ \cline{2-3}
      at the target point & Momentum [GeV/$c$] & 1.10--1.75 \\ \cline{2-3}
                          & Angle [deg] & 0--25 \\ \hline 
      Process & \multicolumn{2}{|c|}{Electromagnetic: ON} \\
              & \multicolumn{2}{|c|}{Hadronic: ON} \\ \hline
      Detector  &\multicolumn{2}{|c|}{TOF (2$^{t}$~cm plastic scintillator)} \\ 
      Materials &\multicolumn{2}{|c|}{WC (pure water + acrylic)} \\ \hline 
      Assumed MsNPE  & \multicolumn{2}{|c|}{200} \\
      in WC & \multicolumn{2}{|c|}{($\beta=1$ particle)}\\
      \hline \hline
    \end{tabular}
  \end{center}
\end{table}


\subsection{Analysis and Results}
\begin{figure}[!htbp]
  \begin{center}
    \includegraphics[width=8.6cm]{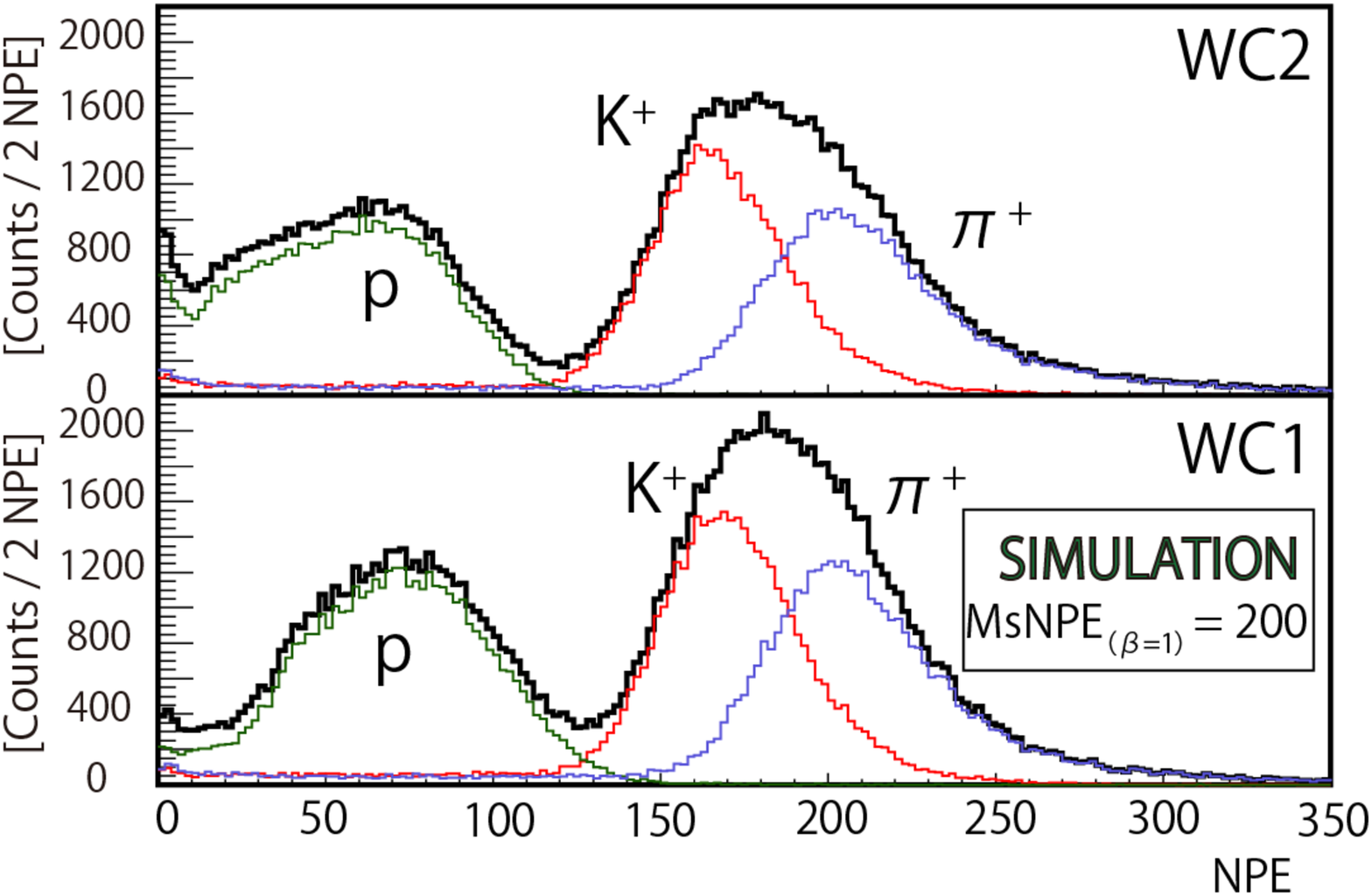}
    \caption{NPE distributions for protons, $K^{+}$, and $\pi^{+}$s 
      in the first and second layers (WC1, WC2) 
      of water {\cerenkov} detectors.}
    \label{fig:npe200}
  \end{center}
\end{figure}

\begin{figure}[!htbp]
  \begin{center}
    \includegraphics[width=8.6cm]{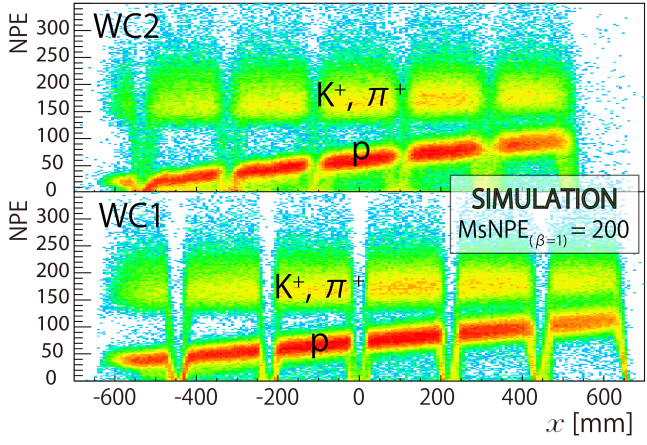}
    \caption{Distributions of NPE vs. $x$-position 
      at the reference plane for WC1 and WC2 in the Monte Carlo simulation.
      The inefficient regions in the layers are designed to be covered by each other.
      Assumed MsNPE in the water {\cerenkov} detector
      for a $\beta=1$ particle is 200.}
    \label{fig:x_vs_npe}
  \end{center}
\end{figure}

Figure~\ref{fig:npe200} shows the 
NPE distributions in the first (WC1) and second (WC2) layers 
of water {\cerenkov} detectors for protons, $K^{+}$s, and $\pi^{+}$s
with the same number of events for each kind of particle.
The NPE distributions in WC2
are broader than those in WC1 owing to secondary particles 
that originate from some reactions in the materials and $K^{+}$ decay.
Two-dimensional plots of NPE vs. horizontal ($x$) position at the 
reference plane\footnote{3~cm upstream of the front surface of the water {\cerenkov} detector.} 
for WC1 and WC2 are shown in Fig.~\ref{fig:x_vs_npe}.
The inefficient region in each layer is covered by another one.

In the experiment, protons will be 
suppressed on-line by using a logical condition as follows: 
\begin{eqnarray}
  WC1 \oplus WC2.
\end{eqnarray}
$WC1$ and $WC2$ indicate that 
an incident particle serves an NPE that 
is above a certain threshold in each layer 
of the water {\cerenkov} detector.
In order to estimate on-line proton-rejection efficiency, 
the following event-selection condition was applied:
\begin{eqnarray}
  WC1^{sim} \oplus WC2^{sim} \label{eq:wc_cut_sim},
\end{eqnarray}
where 
\begin{eqnarray}
  WC1^{sim} = \sum^{6}_{i=1} \Bigl( NPE_{(WC1)}^{i} > x_{1}^{i} \Bigr), \label{eq:wc_cut1} \\
  WC2^{sim} = \sum^{6}_{i=1} \Bigl( NPE_{(WC2)}^{i} > x_{2}^{i} \Bigr). \label{eq:wc_cut2}
\end{eqnarray}
The superscript $i$ represents the segment number in each layer.
The selection thresholds, $x_{1,2}^{i}$, were determined 
so that $>95\%$ $K^{+}$s survive in each segment.



\begin{figure}[!htbp]
  \begin{center}
    \includegraphics[width=8.6cm]{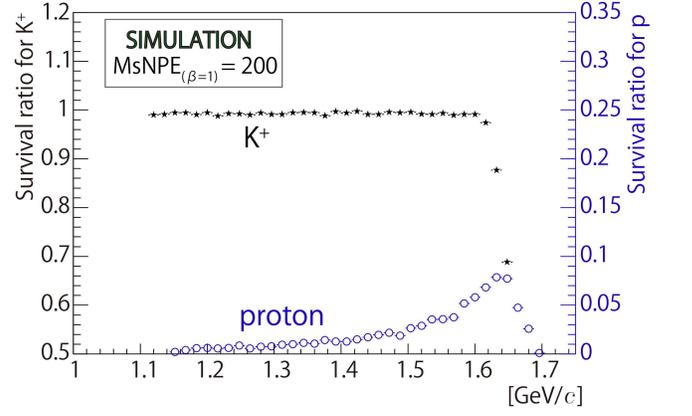}
    \caption{Survival ratios for protons and $K^{+}$s ($SR_{(p,K^{+})}$)
      as a function of particle momentum at the particle generation point.}
    \label{fig:npe_sim}
  \end{center}
\end{figure}

Survival ratios for protons and $K^{+}$s ($SR_{(p,K^{+})}$) are
defined as follows:
\begin{eqnarray}
  SR_{(p,K^{+})} = \frac{N_{(p,K^{+})}}{N^{Ref}_{(p,K^{+})}}, \label{eq:suv_ratio}
\end{eqnarray}
where $N_{(p,K^{+})}$ is 
the number of events with the event selection of Eq.~(\ref{eq:wc_cut_sim}) 
and $N^{Ref}_{(p,K^{+})}$ is that without the event selection at the reference plane.
Fig.~\ref{fig:npe_sim} shows $SR_{(p,K^{+})}$ as a function of 
particle momentum at the particle generation point.
The figure shows that the water {\cerenkov} detector 
is expected to achieve a proton-rejection efficiency $>90\%$ 
while maintaining a survival ratio of $>95\%$ for $K^{+}$s in 
the whole S-2S momentum acceptance.
We found that the performance satisfies our requirements 
for $\Xi^{-}$ hypernuclear spectroscopy at J-PARC.

\section{Summary}
We have developed a water {\cerenkov} detector 
for $\Xi^{-}$ hypernuclear spectroscopy 
by using the new magnetic spectrometer S-2S 
with the {\kk} reaction at the $K^{-}$ momentum of 1.8~GeV/$c$.
The S-2S, which possesses high momentum resolution 
($\Delta p / p \simeq 5.0\times10^{-4}$ in FWHM)
for the momentum region from 1.2 to 1.6~GeV/$c$,
is being constructed for the detection of scattered $K^{+}$.
Protons and $\pi^{+}$s are expected to be detected in S-2S
as major background sources.
In order to suppress protons on-line, 
a water ($n=1.33$) {\cerenkov} detector is 
designed to be installed in S-2S. 

We have constructed prototypes of 
the water {\cerenkov} detector and tested them 
by irradiating positron beams and cosmic-rays. 
The latest prototype was able to achieve 
${\rm MsNPE} \geq 200$ for $\beta=1$ charged particles 
with long-term stability up to 100 days.
A Monte Carlo simulation, 
taking into account information 
obtained from the cosmic-ray test such 
as the MsNPE dependence on the incident particle 
position {\it etc.}, was performed to estimate 
the on-line proton-rejection efficiency. 
Consequently, it is estimated that the developed 
water {\cerenkov} detector is able to achieve a proton-rejection efficiency of $>90\%$ 
while maintaining a $K^{+}$ survival ratio $>95\%$ 
in the whole S-2S momentum acceptance. 
The achieved performance satisfies our requirements 
to perform $\Xi^{-}$ hypernuclear spectroscopy 
using the {\kk} reaction at J-PARC.

\section*{Acknowledgment}
We would like to thank Aquarium Suiso-Kobo company
for their technical support in the prototype design and construction.
We would also like to thank Tohoku ELPH (ELPH experiment $\#2783$) 
and J-PARC for providing us opportunities for test experiments 
using beams.
This work was supported by JSPS KAKENHI Grant Numbers 23000003, 13J01075.  


\end{document}